\documentclass{jfm}
\usepackage{graphicx}
\usepackage{epstopdf, epsfig}
\usepackage{yhmath}
\usepackage{color}
\usepackage{datetime}
\usepackage{url}

\usepackage{boondox-cal}

\shorttitle{Wall-sheared thermal convection}
\shortauthor{A. Xu, B.-R. Xu and H.-D. Xi}

\title{Wall-sheared thermal convection: heat transfer enhancement and turbulence relaminarization}

\author{
Ao Xu\aff{1,2},
Ben-Rui Xu\aff{1}
\and Heng-Dong Xi\aff{1,2}
\corresp{\email{hengdongxi@nwpu.edu.cn}}
}

\affiliation{
\aff{1}School of Aeronautics, Northwestern Polytechnical University, Xi'an 710072, PR China
\aff{2}Institute of Extreme Mechanics, Northwestern Polytechnical University, Xi'an 710072, PR China}

\begin{document}
\maketitle

\begin{abstract}
We studied the flow organization and heat transfer properties in two-dimensional and three-dimensional Rayleigh-B\'enard cells that are imposed with different types of wall shear.
The external wall shear is added with the motivation of manipulating flow mode to control heat transfer efficiency.
We imposed three types of wall shear that may facilitate the single-roll, the horizontally stacked double-roll, and the vertically stacked double-roll flow modes, respectively.
Direct numerical simulations are performed for fixed Rayleigh number $Ra = 10^{8}$ and fixed Prandtl number $Pr = 5.3$, while the wall-shear Reynolds number ($Re_{w}$) is in the range $60 \le Re_{w} \le 6000$.
Generally, we found enhanced heat transfer efficiency and global flow strength with the increase of $Re_{w}$.
However, even with the same magnitude of global flow strength, the heat transfer efficiency varies significantly when the cells are under different types of wall shear.
An interesting finding is that by increasing the wall-shear strength, the thermal turbulence is relaminarized, and more surprisingly, the heat transfer efficiency in the laminar state is higher than that in the turbulent state.
We found that the enhanced heat transfer efficiency at the laminar regime is due to the formation of more stable and stronger convection channels.
We propose that the origin of thermal turbulence laminarization is the reduced amount of thermal plumes.
Because plumes are mainly responsible for turbulent kinetic energy production, when the detached plumes are swept away by the wall shear, the reduced number of plumes leads to weaker turbulent kinetic energy production.
We also quantify the efficiency of facilitating heat transport via external shearing, and find that for larger $Re_{w}$, the enhanced heat transfer efficiency comes at a price of a larger expenditure of mechanical energy.
\footnote{
This article may be downloaded for personal use only.
Any other use requires prior permission of the author and Cambridge University Press.
This article appeared in Xu \emph{et al.}, J. Fluid Mech. \textbf{960}, A2 (2023) and may be found at \url{https://doi.org/10.1017/jfm.2023.173}.
}
\end{abstract}


\section{Introduction}\label{sec:Introduction}

Thermal convection occurs ubiquitously in nature and has wide applications in industry.
A paradigm for the study of thermal convection is the Rayleigh-B\'enard (RB) convection, which is a fluid layer heated from the bottom and cooled from the top \citep{ahlers2009heat,lohse2010small,chilla2012new,xia2013current}.
The control parameters of the canonical RB system include the Rayleigh number ($Ra$, defined later in the paper) that describes the strength of the buoyancy force relative to the thermal and viscous dissipative effects, and the Prandtl number ($Pr$) that represents the thermophysical fluid properties.
One of the response parameters of the RB system is the Nusselt number ($Nu$), which characterizes the global heat transfer efficiency.
Various approaches have been designed to enhance the heat transfer efficiency of the convection cells,
such as adding roughness to the walls \citep{ciliberto1999random,wagner2015heat,jiang2018controlling,rusaouen2018thermal,zhu2019scaling},
introducing vibration forcing \citep{wang2020vibration,yang2020periodically},
adding a dispersed phase of particles or bubbles \citep{lakkaraju2013heat,guzman2016heat,gvozdic2018experimental,wang2019self,yang2022dynamic},
confinement \citep{huang2013confinement,chong2017confined,zhang2022exploring},
rotation \citep{zhong2009prandtl,stevens2009transitions,stevens2013heat,yang2020rotation},
and the addition of passive barriers \citep{liu2020heat}.

\cite{roche2002prandtl} and  \cite{chilla2004long} conjectured that the internal flow structure is correlated with global heat transfer.
\cite{sun2005azimuthal} compared the $Nu$ values in a leveled cell and a tilted cell; correspondingly, the large-scale circulation (LSC) plane sweeps azimuthally or is locked in a particular orientation.
They showed that $Nu$ is larger in the levelled cell, indicating that different flow structures can result in different values of $Nu$.
\cite{xi2008flow} observed both the single-roll and double-roll flow structures in the LSC.
They examined the average $Nu$ corresponding to a particular flow structure, and found that the single-roll flow structure is more efficient for heat transfer.
\cite{weiss2011turbulent} further confirmed the occurrence of a double-roll structure in the LSC, and the higher heat transfer efficiency of the single-roll state.
\cite{van2011connecting,van2012flow} showed numerically that the coexistence of different turbulent structures also exists in simple two-dimensional RB cells with various cell aspect ratios.
They also studied the effect of various velocity boundary conditions (i.e. no-slip, stress-free and periodic boundary conditions) on the heat transfer and flow topology \citep{van2014effect},
and they showed that either the roll-like or the zonal flow can appear under different velocity boundary conditions.
Adopting Fourier mode decompositions, \cite{xi2016higher} presented direct evidence that the first Fourier mode is more efficient for heat transfer in a cylindrical cell.
\cite{xu2020correlation} analysed the coherent flow structure in two-dimensional square convection cells.
Results from both Fourier mode decomposition and proper orthogonal decomposition indicate that the single-roll flow mode and the horizontally stacked double-roll mode are efficient for heat transfer on average;
in contrast, the vertically stacked double-roll mode is inefficient for heat transfer on average.
A natural question arises on how to manipulate flow mode to control heat transfer efficiency.

In this work, we impose various types of wall shear to control the internal flow mode, which  leads further to modification of heat transfer.
Previously, \cite{blass2020flow,blass2021effect} added a Couette-type shear (i.e. the top and bottom walls move in opposite directions with constant speed $u_{w}$) to the RB system as an attempt to trigger the transition to the ultimate convection regime \citep{kraichnan1962turbulent}.
With the increasing wall-shear strength, they observed the variation of flow states from a buoyancy-dominated regime to a shear-dominated regime.
In the buoyancy-dominated regime, the flow structure is similar to that in the canonical RB convection; in the transitional regime, the rolls are increasingly elongated with increasing shear; in the shear-dominated regime, there are large-scale meandering rolls.
\cite{jin2022shear} further added the Couette-type shear to convection cells that have rough walls,
and the moving rough plates introduce an external shear to strengthen the LSC.
As a result, the interactions between the LSC and secondary flows within cavities are increased, and more thermal plumes are triggered.
In this work, our motivation of imposing wall shear is to facilitate various flow modes (i.e. the single-roll, the horizontally stacked double-roll, and the vertically stacked double-roll modes) in the convection cell to further control heat transfer efficiency.
Specifically, we will add the $(m,n)$ type of wall shear to the RB system, and such types of wall shear are expected to facilitate $m$ rolls in the horizontal direction and $n$ rolls in the vertical direction.
The use of shear-modulated boundary conditions leads essentially to mixed convection, which has received considerable attention due to its importance in many engineering applications, such as cooling of electronic devices, coating, and float glass production \citep{hunt1991industrial,shankar2000fluid}.
The rest of this paper is organized as follows.
In section 2, we present numerical details for the simulations.
In section 3, general flow and heat transfer features are presented, and heat transfer enhancement under various types of wall shear is reported.
An interesting finding is thermal turbulence relaminarization under the imposed wall shear, and we then discuss the possible mechanism behind it.
In addition, we quantify the efficiency of facilitating heat transport via external shearing.
In section 4, the main findings of the present work are summarized.

\section{Numerical method}

\subsection{Direct numerical simulation of incompressible thermal convection}

We consider incompressible thermal convection under the Boussinesq approximation.
The temperature is treated as an active scalar, and its influence on the velocity field is realized through the buoyancy term;
all the transport coefficients are assumed to be constants.
The governing equations can be written as
\begin{equation}
\nabla \cdot \mathbf{u} = 0 \label{Eq.divU}
\end{equation}
\begin{equation}
\frac{\partial \mathbf{u}}{\partial t}+\mathbf{u}\cdot \nabla \mathbf{u}=-\frac{1}{\rho_{0}}\nabla P+\nu \nabla^{2}\mathbf{u}+g\beta(T-T_{0})\hat{\mathbf{y}} \label{Eq.momentum}
\end{equation}
\begin{equation}
\frac{\partial T}{\partial t}+\mathbf{u}\cdot\nabla T=\alpha \nabla^{2}T \label{Eq.temperature}
\end{equation}
where $\mathbf{u}$  is the fluid velocity, and $P$ and $T$  are the pressure and temperature of the fluid, respectively.
Here, $\beta$, $\nu$ and  $\alpha$ are the thermal expansion coefficient, kinematic viscosity and thermal diffusivity, respectively.
The zero subscripts refer to the reference values;
$g$ is the gravity acceleration value, and $\hat{\mathbf{y}}$ is the unit vector parallel to the gravity.
Using the non-dimensional group
\begin{equation}
    \begin{split}
& \mathbf{x}^{*}=\mathbf{x}/H, \ \ \ t^{*}=t/\sqrt{H/(g\beta \Delta_{T})}, \ \ \ \mathbf{u}^{*}=\mathbf{u}/\sqrt{g \beta \Delta_{T}H}, \\
& P^{*}=P/(\rho_{0}g\beta \Delta_{T}H), \ \ \ T^{*}=(T-T_{0})/\Delta_{T} \\
    \end{split}
\end{equation}
(\ref{Eq.divU})-(\ref{Eq.temperature}) can be rewritten in dimensionless form as
\begin{equation}
\nabla \cdot \mathbf{u}^{*} = 0 \label{Eq.divU-dim}
\end{equation}
\begin{equation}
\frac{\partial \mathbf{u}^{*}}{\partial t^{*}}+\mathbf{u}^{*}\cdot \nabla \mathbf{u}^{*}
=-\nabla P^{*}+\sqrt{\frac{Pr}{Ra}} \nabla^{2}\mathbf{u}^{*}+T^{*}\hat{\mathbf{y}} \label{Eq.momentum-dim}
\end{equation}
\begin{equation}
\frac{\partial T^{*}}{\partial t^{*}}+\mathbf{u}^{*}\cdot\nabla T^{*}=\sqrt{\frac{1}{Pr Ra}} \nabla^{2}T^{*} \label{Eq.temperature-dim}
\end{equation}
Here, $H$  is the cell height, and $\Delta_{T}$ is the temperature difference between heating and cooling walls.
In the following, for convenience, we will drop the superscript star ($*$) to denote a dimensionless variable.
The dimensionless parameters of the Rayleigh number ($Ra$), the Prandtl number ($Pr$) and the cell aspect ratio ($\Gamma_{\parallel}$ in the plane parallel to the LSC plane, and $\Gamma_{\perp}$ in the plane perpendicular to the LSC) are defined as
\begin{equation}
Ra=\frac{g\beta \Delta_{T}H^{3}}{\nu \alpha}, \ \ \ Pr=\frac{\nu}{\alpha}, \ \ \ \Gamma_{\parallel}=\frac{L}{H}, \ \ \ \Gamma_{\perp}=\frac{W}{H}
\end{equation}
where $L$ is cell length and $W$ is cell width.

We adopt the spectral element method \citep{patera1984spectral} implemented in the open-source Nek5000 solver (version v19.0) as the numerical tool for the direct numerical simulation.
In the Nek5000 solver, the effective grid number equals the product of spectral element number and polynomial order.
We set the spectral elements for the velocity with polynomial order $N$, and the spectral elements for the pressure with polynomial order $N-2$ (to avoid spurious pressure modes).
Similar to previous turbulent flow simulations \citep{kooij2018comparison}, we fix the polynomial order as $N=8$.
The viscous term is treated implicitly with the second-order backward difference scheme, while the convection term and other terms are treated with an explicit second-order extrapolation scheme.
The discretized system is solved with preconditioned conjugate gradient (PCG) iteration, and Jacobi preconditioning is adopted for the linear velocity system.
A pressure correction step follows the solution of the discretized system, which is also solved with PCG iteration; and the linear pressure system is solved by the multilevel overlapping Schwarz method.
As for the energy equation (i.e. temperature governed by a convection-diffusion type equation), the transient term is treated implicitly with the second-order backward difference scheme, and the convection term is treated with an explicit second-order extrapolation scheme.
For the Navier-Stokes and convection-diffusion equations, the temporal derivative applies a Courant-Friedrichs-Lewy constraint $\max(|\mathbf{u}|\Delta_{t}/\Delta_{x})\approx 0.5$.
More numerical details of the spectral element method and validation of the Nek5000 solver can be found in \citep{fischer1997overlapping,fischer2002spectral,deville2002high,kooij2018comparison}.
To verify the results obtained from the Nek5000 solver, we also performed a set of simulations at wall-shear Reynolds number ($Re_{w}$, defined later in the paper) 100 using an in-house solver based on the lattice Boltzmann method \citep{xu2017accelerated,xu2019lattice,xu2023multi}.
The results from the open-source Nek5000 solver and the in-house lattice Boltzmann solver are shown to be consistent.

\subsection{Simulation settings}

\begin{figure}
  \centerline{\includegraphics[width=0.86\textwidth]{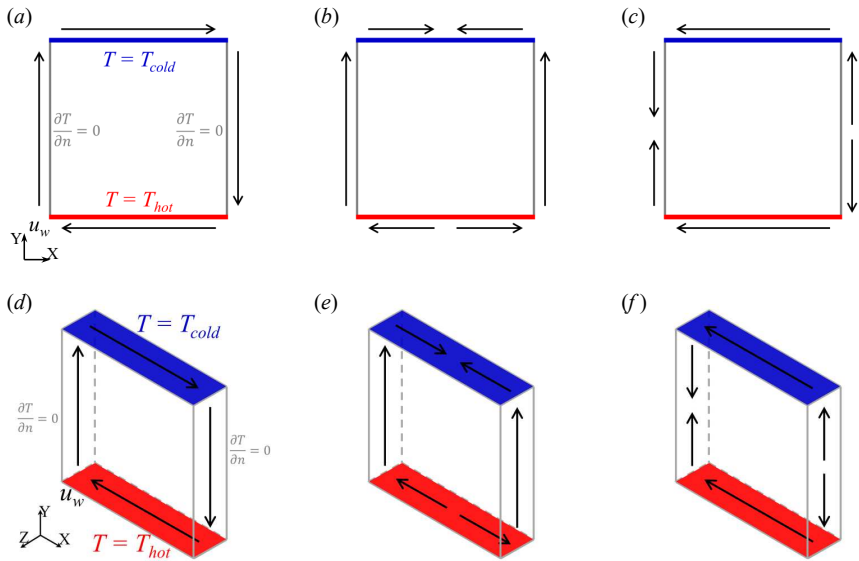}}
  \caption{Schematic illustration of the shear convection cells in (\textit{a-c}) two-dimensions and (\textit{d-f}) three-dimensions, for (\textit{a}, \textit{d}) the (1, 1) type wall shear, (\textit{b}, \textit{e}) the (2, 1) type wall shear, and (\textit{c}, \textit{f}) the (1, 2) type wall shear boundary conditions.}
\label{fig:demo}
\end{figure}

As illustrated in figure \ref{fig:demo}, the dimensions $H$, $L$ and $W$ correspond to  $y$, $x$ and $z$ in Cartesian coordinates.
The top and bottom of the horizontal walls are kept at constant low and high temperatures $T_{cold}$ and $T_{hot}$, respectively, while the vertical sidewalls are adiabatic.
For the velocity at the walls, we designed the $(m, n)$ type wall shear to facilitate the flow structure with $m$ rolls in the $x$-direction and $n$ rolls in the $y$-direction.
Specifically, we consider three types of wall-shear boundary conditions:
the (1, 1) type wall shear that may facilitate the single-roll flow mode (see figures \ref{fig:demo}\textit{a},\textit{d});
the (2, 1) type wall shear that may facilitate the horizontally stacked double-roll mode (see figures \ref{fig:demo}\textit{b},\textit{e});
and the (1, 2) type wall shear that may facilitate the vertically stacked double-roll mode (see figures \ref{fig:demo}\textit{c},\textit{f}).
Under the (1, 1) type wall shear, the velocity boundary conditions are:
(i) at $0 \le x \le L$ and $y=0$, we have $\mathbf{u}=(-u_{w}, 0, 0)$;
(ii) at $0 \le x \le L$ and $y=H$, we have $\mathbf{u}=(u_{w}, 0, 0)$;
(iii) at $x = 0$ and $0 \le y \le H$, we have $\mathbf{u}=(0, u_{w}, 0)$;
(iv) at $x=L$ and $0 \le y \le H$, we have $\mathbf{u}=(0, -u_{w}, 0)$.
Similar mathematical formulations for the velocity boundary conditions under the (2, 1) type and the (1, 2) type wall shear can be written easily (not present here for clarity).
When an external wall shear is introduced, an additional control parameter of wall-shear Reynolds number ($Re_{w}=Hu_{w}/\nu$) is needed.
Here, $u_{w}$  is the wall-shear velocity.
Simulation results are provided for fixed Rayleigh number $Ra = 10^{8}$, fixed Prandtl number $Pr = 5.3$ [corresponds to the working fluids of water at 31$^{\circ}$C \citep{zhang2017statistics}] and fixed aspect ratio $\Gamma_{\parallel}=1$.
In the three-dimensional (3-D) cases, we consider aspect ratios $\Gamma_{\perp}=1/8$ and 1/4 such that the LSC is confined in the $x-y$ plane, enabling easy manipulation of the flow mode via wall shear.
The wall-shear Reynolds number is in the range $60 \le Re_{w} \le 6000$ for two-dimensional (2-D) cases, and $Re_{w}=100$ and 3000 for 3-D cases.

In the simulation, after the initial transient stage, we run at least 5000 $t_{f}$ for 2-D cases and 800 $t_{f}$ for 3-D cases to obtain the statistics.
Here, $t_{f}$ denotes free-fall time units: $t_{f}=\sqrt{H/(g\beta \Delta_{T})}$.
We check whether the grid spacing $\Delta_{g}$ and time interval $\Delta_{t}$ are properly resolved by comparing them with the Kolmogorov and Batchelor scales.
The Kolmogorov length scale can be estimated as $\eta_{K}=(\nu^{3}/\langle \varepsilon_{u} \rangle)^{1/4}$,
the Batchelor length scale can be estimated as $\eta_{B}=\eta_{K}Pr^{-1/2}$ \citep{batchelor1959small,silano2010numerical},
and the Kolmogorov time scale can be estimated as $\tau_{\eta}=\sqrt{\nu/\langle \varepsilon \rangle}$.
In the canonical RB convection, we adopted spectral elements of $64 \times 64$ for 2-D cases, $32 \times 32 \times 5$ for 3-D cases with $\Gamma_{\perp}=1/8$, and $32 \times 32 \times 9$ for 3-D cases with $\Gamma_{\perp}=1/4$;
the corresponding effective grid numbers are listed in table \ref{table:resolution}.
In the wall-sheared thermal convection, we adopted a finer distributed spectral element of $96 \times 96$ for 2-D cases, $44 \times 44 \times 7$ for 3-D cases with $\Gamma_{\perp}=1/8$, and $44 \times 44 \times 13$ for 3-D cases with $\Gamma_{\perp}=1/4$.
We estimate the global kinetic energy dissipation rate as $\langle \varepsilon_{u} \rangle=RaPr^{-2}(Nu-1)\nu^{3}/H^{4}$ in the canonical RB convection \citep{shraiman1990heat},
and $\langle \varepsilon_{u} \rangle = \sqrt{Pr/Ra\langle (\partial_{j}u_{i}')\rangle_{V,t}} $ in the wall-sheared convection \citep{pope2000turbulent}.
Here, the subscripts $i$ and $j$ are dummy indices, and $\langle \cdot \rangle_{V,t}$  denotes the spatial and temporal average.
As shown in table \ref{table:resolution}, the maximum grid spacing $(\Delta_{g})_{\max}$ is less than (or comparable) to the Kolmogorov and Batchelor length scales for 2-D cases (or 3-D cases);
the maximum time interval $(\Delta_{t})_{\max}$ is far less than the Kolmogorov time scale for all the cases.
Thus adequate spatial and temporal resolution is guaranteed.
Each simulation was conducted with 48 message passing interface processes on an in-house cluster that required around 12 000 core hours for 2-D cases and 50 000 core hours for 3-D cases.

\begin{table}
\centering
\begin{tabular}{cccccccccccc}

Wall shear type & $Re_{w}$ &   $\Gamma_{\perp}$   & Effective grid number & $(\Delta_{g})_{\max}/\eta_{K}$ & $(\Delta_{g})_{\max}/\eta_{B}$ & $(\Delta_{t})_{\max}/\tau_{\eta}$  \\
\\
   -          &    0     &  -    & $512 \times 512$             & 0.26         & 0.60                           & 0.0019 \\
              &    0     &  1/8  & $256 \times 256 \times 40$   & 0.51         & 1.17                           & 0.0068 \\
              &    0     &  1/4  & $256 \times 256 \times 72$   & 0.52         & 1.17                           & 0.0054 \\
\\
(1, 1) type   &    60    &  -    & $768 \times 768$             & 0.13         & 0.30                           & 0.0009 \\
              &   100    &  -    & $768 \times 768$             & 0.13         & 0.30                           & 0.0008 \\
              &   200    &  -    & $768 \times 768$             & 0.13         & 0.30                           & 0.0008 \\
              &   500    &  -    & $768 \times 768$             & 0.12         & 0.28                           & 0.0007 \\
              &   800    &  -    & $768 \times 768$             & 0.11         & 0.24                           & 0.0005 \\
              &   100    &  1/8  & $352 \times 352 \times 56$   & 0.34         & 0.77                           & 0.0046 \\
              &   100    &  1/4  & $352 \times 352 \times 104$  & 0.35         & 0.80                           & 0.0032 \\
\\
(2, 1) type   &    60    &  -    & $768 \times 768$             & 0.15         & 0.35                           & 0.0012 \\
              &   100    &  -    & $768 \times 768$             & 0.14         & 0.33                           & 0.0010 \\
              &   200    &  -    & $768 \times 768$             & 0.14         & 0.32                           & 0.0009 \\
              &   500    &  -    & $768 \times 768$             & 0.13         & 0.31                           & 0.0006 \\
              &   100    &  1/8  & $352 \times 352 \times 56$   & 0.32         & 0.74                           & 0.0041 \\
              &   100    &  1/4  & $352 \times 352 \times 104$  & 0.34         & 0.78                           & 0.0032 \\
              &  3000    &  1/4  & $352 \times 352 \times 104$  & 0.55         & 1.26                           & 0.0017 \\
\\
(1, 2) type   &    60    &  -    & $768 \times 768$             & 0.17         & 0.39                           & 0.0012 \\
              &   100    &  -    & $768 \times 768$             & 0.14         & 0.33                           & 0.0009 \\
              &   200    &  -    & $768 \times 768$             & 0.13         & 0.30                           & 0.0007 \\
              &   500    &  -    & $768 \times 768$             & 0.12         & 0.27                           & 0.0007 \\
              &   800    &  -    & $768 \times 768$             & 0.11         & 0.25                           & 0.0004 \\
              &  1000    &  -    & $768 \times 768$             & 0.10         & 0.23                           & 0.0003 \\
              &   100    &  1/8  & $352 \times 352 \times 56$   & 0.34         & 0.79                           & 0.0038 \\
              &   100    &  1/4  & $352 \times 352 \times 104$  & 0.38         & 0.87                           & 0.0039 \\
              &  3000    &  1/4  & $352 \times 352 \times 104$  & 0.53         & 1.22                           & 0.0016 \\
\end{tabular}
\caption{A \emph{posteriori} check of spatial and temporal resolutions of the simulations.
The columns from left to right indicate the following: imposed wall shear type ('-' denotes convection without wall shear),
the wall shear Reynolds number $Re_{w}$,
cell aspect ratio $\Gamma_{\perp}$ in the plane perpendicular to the LSC ('-' denotes 2-D cases),
effective grid number (i.e., the product of spectral element number and polynomial order),
the ratio of maximum grid spacing over the Kolmogorov length scale,
the ratio of maximum grid spacing over the Batchelor length scale,
the ratio of maximum time interval over the Kolmogorov time scale.
Note that not all the simulations in this work are listed in the table.}
\label{table:resolution}
\end{table}

\section{Results and discussion}
\subsection{Global flow and heat transfer features}

Typical snapshots of temperature field and flow field under the three types of wall shear are shown in figures \ref{fig:Instant-Flow-Field}, and the corresponding video can be viewed in supplementary movie 1 available at https://doi.org/10.1017/jfm.2023.173.
Here, $Ra$ is fixed as $Ra=10^{8}$ and $Pr$ is fixed as $Pr=5.3$.
At small wall-shear strength $Re_{w} = 100$, the convection is still buoyancy-dominated, and plumes detach from thermal boundary layers and further self-organize into the LSC;
meanwhile, the flow structure in the convection cell is influenced by the imposed wall shear.
For the convenience of comparison, we also provide the flow and heat transfer patterns in the canonical RB convection without wall shear (see Appendix A).
The single-roll flow structure appears under the (1, 1) type wall shear (see figures \ref{fig:Instant-Flow-Field}\textit{a},\textit{g}), whilst the corner rolls are suppressed compared to that without wall shear;
the horizontally stacked double-roll flow structure appears under the (2, 1) type wall shear (see figures \ref{fig:Instant-Flow-Field}\textit{b},\textit{h});
and the vertically stacked double-roll flow structure appears under the (1, 2) type wall shear (see figures \ref{fig:Instant-Flow-Field}\textit{c},\textit{i}).
At large wall-shear strength $Re_{w} = 4000$ in two-dimensions and $Re_{w}=3000$ in three-dimensions, the convection is shear-dominated, and the flow structures inside the convection cell are completely influenced by the external wall shear.
For example, under the (1, 1) type wall shear (see figures \ref{fig:Instant-Flow-Field}\textit{d},\textit{j}), the hot (or cold) fluids near the bottom (or top) wall are swept away by the LSC in the clockwise direction, and rise (or fall) along the left (or right) vertical wall, while the fluids in the bulk region are well-mixed.
Similar observations can be found for the flow structure under the (2, 1) type wall shear (see figures \ref{fig:Instant-Flow-Field}\textit{e},\textit{k}), while the cold fluids also fall along the vertical mid-plane of the cell.
As for the flow structure under the (1, 2) type wall shear, in the 2-D case (see figure \ref{fig:Instant-Flow-Field}\textit{f}), the top and bottom subregions are completed separated without heat transfer between them, acting as a 'thermal barrier' exists at the half-height of the cell;
however, in the 3-D case  (see figure \ref{fig:Instant-Flow-Field}\textit{l}), we did not observe a complete separation of hot and cold fluids.
We infer that the differences in flow structure between 2-D and 3-D configurations are due to the flow state:
in the steady laminar flow (as in the 2-D case), the rising hot fluids and falling cold fluids can remain stable boundaries;
while in the turbulent flow (as in the 3-D case), the hot and cold fluids are more mixed.
We also checked the flow field within the $\Gamma_{\perp}=1/8$ cell, where the flow is in a laminar state, and we indeed found a separation of hot and cold fluids.
\begin{figure}
  \centerline{\includegraphics[width=0.95\textwidth]{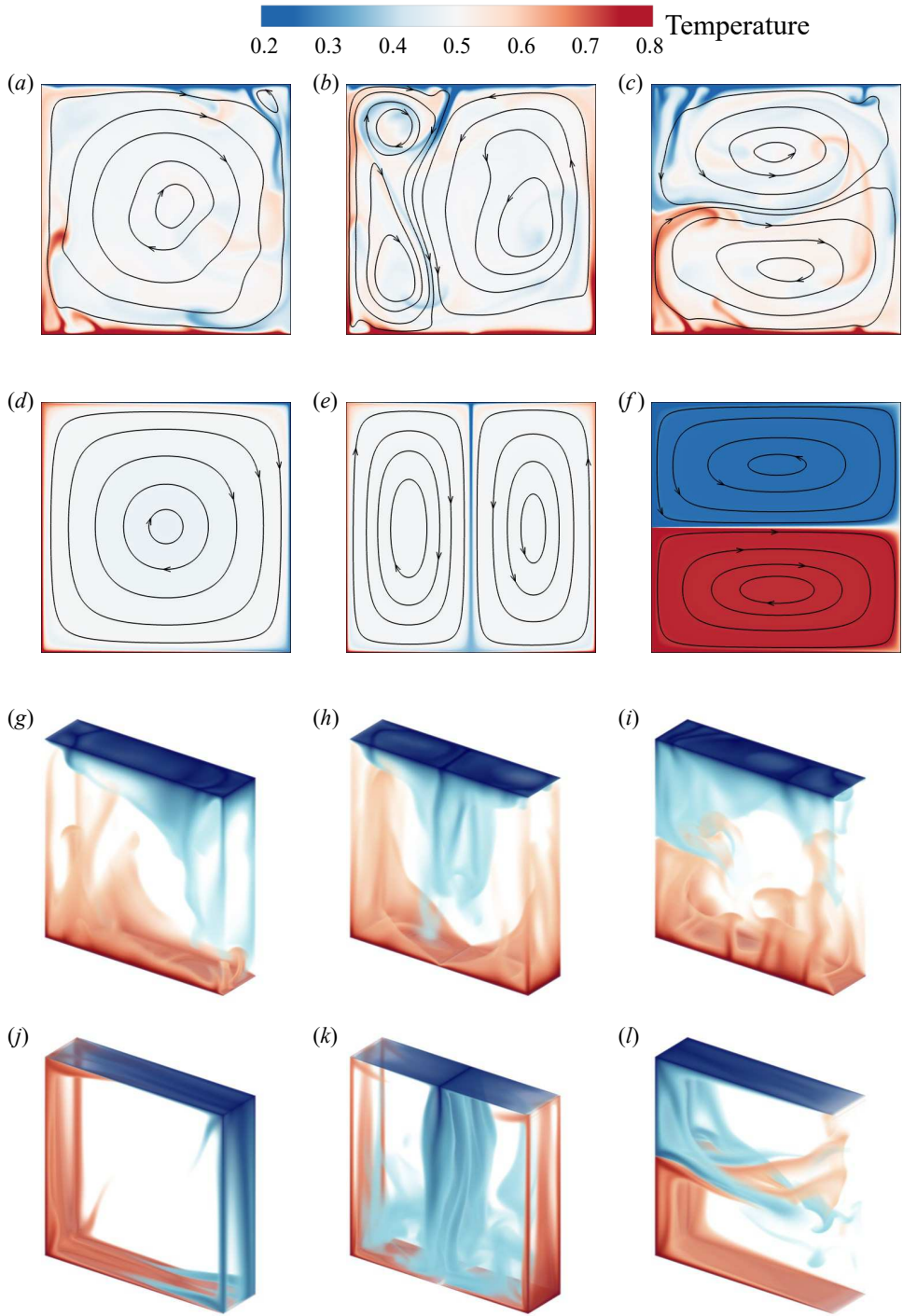}}
  \caption{Typical instantaneous temperature field (contours in two dimensions and volume rendering in three dimensions) and flow field (streamlines in two dimensions) at (\textit{a}-\textit{c}) $Re_{w} = 100$, (\textit{d}-\textit{f}) $Re_{w} = 4000$, (\textit{g}-\textit{i}) $Re_{w} = 100$ and $\Gamma_{\perp}=1/4$, (\textit{j}-\textit{l}) $Re_{w} = 3000$ and $\Gamma_{\perp}=1/4$,
  under (left-column) the (1, 1) type wall shear, (middle-column) the (2, 1) type wall shear, and (right-column) the (1, 2) type wall shear.}
\label{fig:Instant-Flow-Field}
\end{figure}

\begin{figure}
  \centerline{\includegraphics[width=0.8\textwidth]{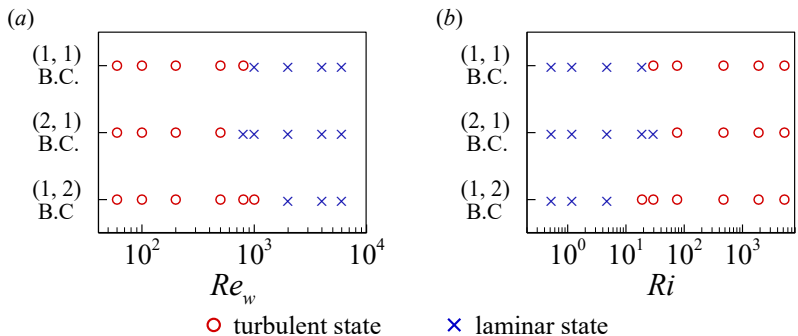}}
  \caption{Phase diagram of the flow states (\textit{a}) at different $Re_{w}$, and (\textit{b}) at different $Ri$, in 2-D cases.}
\label{fig:flowStates}
\end{figure}

With simulations of three different types of wall shear in the range $60 \le Re_{w} \le 6000$, we can obtain the phase diagram of whether the flow is in the turbulent state or laminar state, as shown in figure \ref{fig:flowStates}(\textit{a}) for 2-D cases.
Here, we placed numerical probers in the cell and analysed the time recordings of local velocity and temperature series to determine the flow states \citep{heslot1987transitions,silano2010numerical}.
We determined that the flow is in the laminar state if the time recordings do not vary with time (i.e. steady laminar state) or the power spectral density (PSD) of the time recordings exhibits characteristic peaks (i.e. unsteady laminar state);
otherwise, if the PSD of the time recordings exhibits continuous spectra, then the flow is in the turbulent state.
In Appendix B,  we give examples of temperature series and the corresponding PSD at the location (0.25, 0.5) in the 2-D convection cell under (1, 1) type wall shear.
The phase diagram of the flow states can be understood in terms of competition between buoyancy and shear effects, which can be quantified by the Richardson number as $Ri=Ra/(Re_{w}^{2}Pr)$.
In figure \ref{fig:flowStates}(\textit{b}), we redraw the phase diagram of the flow states at different $Ri$.
For lower $Re_{w}$ (i.e. higher $Ri$ at fixed $Ra$ and $Pr$), the flow is buoyancy-dominated and possesses the key features of turbulent convection;
for higher $Re_{w}$ (i.e. lower $Ri$), the flow is shear-dominated and enters a laminar state.
Turbulent laminarization is counterintuitive and is found in pipe flow by amplifying wall shear \citep{kuhnen2018destabilizing,scarselli2019relaminarising}.
It should also be noted that when $Re_{w}$ increases further, the wall shear would introduce flow instability and the flow would transit to a turbulent state again.
However, our numerical tests show that the flow can remain laminar for a wide range of $Re_{w}$ in 2-D cases;
a further transition to shear turbulence may occur at a much higher $Re_{w}$.
We also found that the shear instability is prominent in 3-D cases, particularly when $\Gamma_{\perp}$ is larger, thus the flow remains laminar in a smaller range of wall-shear Reynolds number.
For example, at $Re_{w}=3000$, the flow is laminar in convection cells with $\Gamma_{\perp}=1/8$ under all three types of wall shear,
while the flow is laminar only under the (1, 1) type wall shear in the $\Gamma_{\perp}=1/4$ cell.

\begin{figure}
  \centerline{\includegraphics[width=0.84\textwidth]{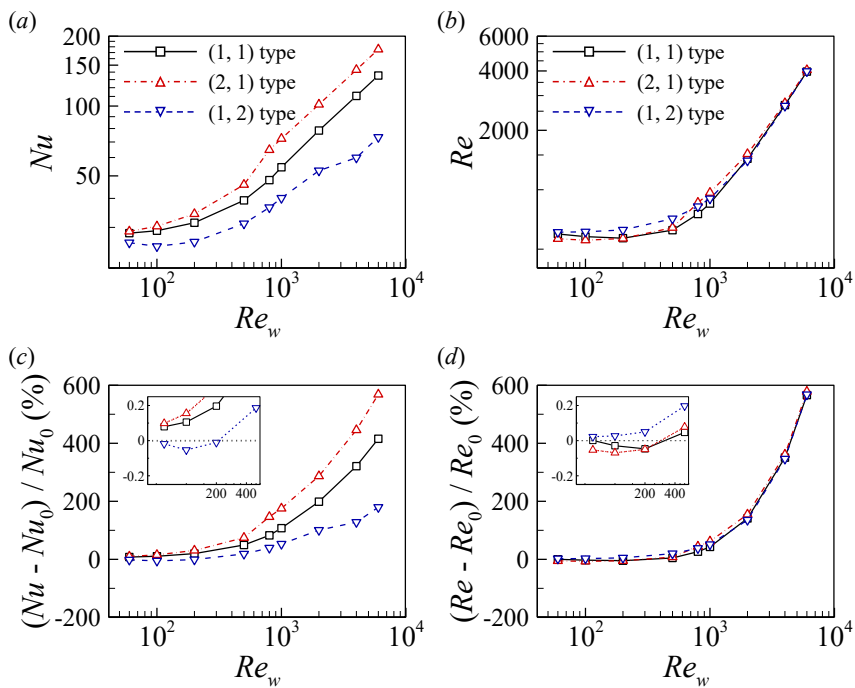}}
  \caption{(\textit{a}) Nusselt number, (\textit{b}) Reynolds number, (\textit{c}) values of $Nu/Nu_{0}-1$, and ($\textit{d}$) values of $Re/Re_{0}-1$, as functions of $Re_{w}$ for various types of wall shear in the 2-D cases.
  Here, $Nu_{0}$ and $Re_{0}$ are the Nusselt and Reynolds numbers in the absence of wall shear, respectively.
  The insets magnify $Re_{w}$ in the range $100 \le Re_{w} \le 500$.}
\label{fig:ReNu}
\end{figure}

We then examine the global response parameters of Nusselt number ($Nu$) and Reynolds number ($Re$) on the control parameter $Re_{w}$.
Here, the heat transfer efficiency is calculated as  $Nu=\sqrt{Ra Pr}\langle vT \rangle_{V,t}+1$, and the global flow strength is calculated as  $Re=\sqrt{\langle \|\mathbf{u}\|^{2} \rangle_{V,t}}H/\nu$.
The measured $Nu$ and $Re$ as functions of $Re_{w}$ for various types of wall shear in 2-D cells are shown in figures \ref{fig:ReNu}(\textit{a}) and \ref{fig:ReNu}(\textit{b}), respectively.
Generally, with the increase of $Re_{w}$, we can observe enhanced heat transfer efficiency and global flow strength for all three types of wall shear.
However, at $Re_{w} \le 200$ for the (1, 2) type wall shear,
the flow structure gradually changes from an LSC that spans the whole cell to the vertically stacked double-roll mode, leading to a decreased $Nu$ value \citep{xu2020correlation}.
To clearly visualize the relative changes of $Nu$ and $Re$ after imposing the wall shear, we further plot $(Nu-Nu_{0})/Nu_{0}$ and $(Re-Re_{0})/Re_{0}$ as functions of $Re_{w}$ in figures \ref{fig:ReNu}(\textit{c}) and \ref{fig:ReNu}(\textit{d}), respectively.
Here, $Nu_{0}$ and $Re_{0}$ are the Nusselt and Reynolds numbers in the absence of wall shear, respectively.
Among the three types of wall shear, at the same $Re_{w}$, the (2, 1) type wall shear results in the largest magnitude of heat transfer efficiency up to 568\%; and the (1, 2) type wall shear results in the smallest one, approximately 179\%.
The trend is consistent with our expectation that facilitating the horizontally stacked double-roll flow modes is efficient for heat transfer, yet facilitating the vertically stacked double-roll is inefficient for heat transfer \citep{xu2020correlation}.
On the other hand, as $Re_{w}$ increases, all three types of wall shear exhibit a similar trend of increasing global flow strength.
The results indicate that in even with the same magnitude of flow strength, the heat transfer efficiency of the convection cell still varies significantly under different types of wall shear.
In addition, we provide tabulated value of Nusselt and Reynolds numbers for 3-D cases in table \ref{table:ReNu}.
We can conclude that heat transfer enhancement can also be found in 3-D configurations.

\begin{table}
\centering
\begin{tabular}{ccccccc}
Wall shear type &   $\Gamma_{\perp}$   & $Re_{w}$ &  $Nu$  & $Re$  & $(Nu-Nu_{0})/Nu_{0}$ & $(Re-Re_{0})/Re_{0}$  \\
\\
-             &   1/8   &   0      & 34.80    & 258.39   & -       & -             \\
              &   1/4   &   0      & 32.38    & 282.32   & -       & -              \\
\\
(1, 1) type   &   1/8   &   100    & 35.18    & 280.65   & 1.1\%    & 8.6\%              \\
              &   1/8   &   3000   & 82.44    & 741.02   & 136.9\%  & 186.8\%              \\
              &   1/4   &   100    & 34.32    & 357.81   & 6.0\%    & 26.7\%             \\
              &   1/4   &   3000   & 86.39    & 854.66   & 166.8\%  & 202.7\%             \\
\\
(2, 1) type   &   1/8   &   100    & 35.86    & 273.78   & 3.0\%    & 6.0\%              \\
              &   1/8   &   3000   & 112.15   & 761.90   & 222.2\%  & 194.9\%              \\
              &   1/4   &   100    & 36.05    & 351.58   & 11.3\%   & 24.5\%              \\
              &   1/4   &   3000   & 113.75   & 910.96   & 251.3\%  & 222.7\%              \\
\\
(1, 2) type   &   1/8   &   100    & 34.62    & 265.11   & -0.5\%   & 2.6\%              \\
              &   1/8   &   3000   & 68.83    & 784.14   & 97.8\%   & 203.5\%              \\
              &   1/4   &   100    & 33.16    & 298.15   & 2.4\%    & 5.6\%              \\
              &   1/4   &   3000   & 67.95    & 935.05   & 109.9\%  & 231.2\%              \\
\end{tabular}
\caption{Heat transfer efficiency and global flow strength in the 3-D cases.
The columns from left to right indicate the following:
imposed wall shear type ('-' denotes convection without wall shear),
cell aspect ratio ($\Gamma_{\perp}$) in the plane perpendicular to the LSC,
wall shear strength ($Re_{w}$),
Nusselt number ($Nu$),
Reynolds number ($Re$),
heat transfer enhancement $(Nu-Nu_{0})/Nu_{0}$,
global flow strength enhancement $(Re-Re_{0})/Re_{0}$.
Here, $Nu_{0}$ and $Re_{0}$ are the Nusselt and Reynolds numbers in the absence of wall shear at the same $\Gamma_{\perp}$, respectively.}
\label{table:ReNu}
\end{table}

\begin{figure}
  \centerline{\includegraphics[width=0.84\textwidth]{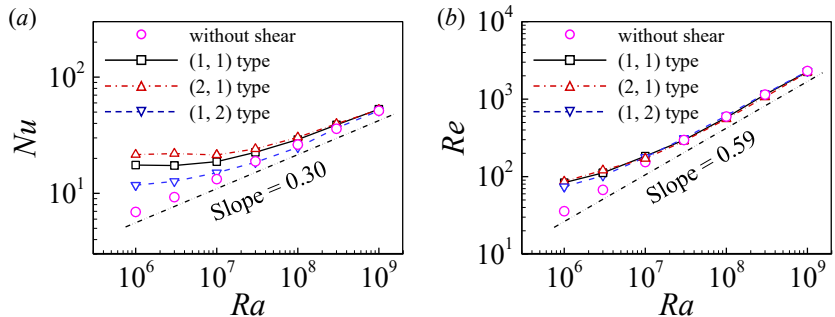}}
  \caption{(\textit{a}) Nusselt number, (\textit{b}) Reynolds number as functions of Rayleigh number for various types of wall shear in the 2-D cases, when the wall-shear Reynolds number is fixed as $Re_{w}=100$.}
\label{fig:scalingLaw}
\end{figure}

Figure \ref{fig:scalingLaw} shows the scaling of the global quantities in 2-D cells, such as $Nu$ and $Re$, on one of the control parameters $Ra$ (for $10^{6} \le Ra \le 10^{9}$), whilst the control parameter $Re_{w}$ is fixed as $Re_{w}=100$, and $Pr$ is fixed as $Pr=5.3$.
We also provide $Nu$ and $Re$ in the canonical RB convection without shear.
Previously, \cite{zhang2017statistics} provided tabulated values of $Nu$ and $Re$ versus $Ra$ at $Pr = 5.3$.
Our simulation results on the canonical RB convection are in good agreement with those reported by \cite{zhang2017statistics}.
The data shown in figure \ref{fig:scalingLaw} indicate that in the buoyancy-dominated regime (i.e. when $Ra$ is larger at fixed $Re_{w}$),
the increase of $Nu$ and $Re$ gradually approaches the power-law relations $Nu \propto Ra^{0.30}$ and $Re \propto Ra^{0.59}$, consistent with previous results reported in the canonical RB convection \citep{ciliberto1996large,van2012flow,huang2016effects,zhang2017statistics,xu2021tristable}.
Overall, the global heat transfer and momentum quantities reveal that the simulated system possesses the key features of turbulent convection in the buoyancy-dominated regime.
In the shear-dominated regime (i.e. when $Ra$ is smaller at fixed $Re_{w}$), the scaling behaviour of $Nu$ and $Re$ with $Ra$ deviates significantly from that of the canonical RB convection,
suggesting that heat transfer and momentum exchange are not governed solely by the boundary layer.

\begin{figure}
  \centerline{\includegraphics[width=0.5\textwidth]{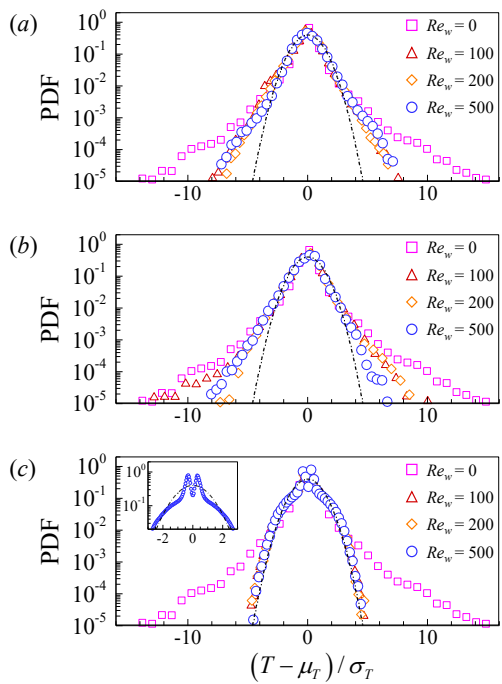}}
  \caption{The probability density functions (p.d.f.s) of the temperature measured in the bulk region of the 2-D cells (i.e. $0.4L \le x \le 0.6L$ and $0.4H \le y \le 0.6H$) under (\textit{a}) the (1, 1) type wall shear, (\textit{b}) the (2, 1) type wall shear, and (\textit{c}) the (1, 2) type wall shear, when the flow is in turbulent state.
  The dot-dashed line represents a Gaussian distribution.
   The inset in (\textit{c}) magnifies the head of the p.d.f. at $Re_{w}=500$.}
\label{fig:pdfT}
\end{figure}

We further investigate quantitatively the influence of different types of wall shear on the temperature distribution.
Figure \ref{fig:pdfT} shows the probability density functions (p.d.f.s) of the normalized temperature $(T-\mu_{T})/\sigma_{T}$ in the bulk region of the 2-D cell (i.e. $0.4L \le x \le 0.6L$ and $0.4H \le y \le 0.6H$), where $\mu_{T}$  and $\sigma_{T}$  are the mean and standard deviation of the temperature.
In the absence of wall shear, the p.d.f.s of temperature in the bulk show a stretched exponential behaviour.
Under the (1, 1) type wall shear, the temperature in the bulk is well-mixed, and the p.d.f.s are symmetric at different $Re_{w}$ (see figure \ref{fig:pdfT}\textit{a}).
With imposed external wall shear, the p.d.f.s at different $Re_{w}$ collapse, and they deviate significantly from that in the absence of wall shear.
The narrowed p.d.f. tails imply that fewer plumes pass through the bulk region and the temperature fluctuation is suppressed.
Under the (2, 1) type wall shear, the p.d.f. is negatively skewed at smaller $Re_{w}$ (see figure \ref{fig:pdfT}\textit{b}), which is due to cold plumes descending through the central region.
However, as $Re_{w}$ increases, the skewness of the temperature p.d.f.s decreases, and their tails become narrower, implying that temperature is better mixed and fewer cold fluids pass through the central region.
Under the (1, 2) type wall shear, the p.d.f.s are symmetric (see figure \ref{fig:pdfT}\textit{c}) due to the top-down symmetry of the convection cell, both hot and cold plumes passing through the central region.
As the strength of the wall shear increases, the heads of the p.d.f.s gradually exhibit a bi-modal shape (e.g. the inset shown in figure \ref{fig:pdfT}\textit{c}), suggesting that the top cold and bottom hot subregions are gradually separated;
meanwhile, all the tails of the p.d.f.s exhibit Gaussian shape, and their profiles collapse for different $Re_{w}$.
The collapse of the p.d.f. indicates a similar flow pattern in the bulk region because the functional form of the temperature p.d.f. is determined by the coherence of plumes \citep{solomon1990sheared,xia1997turbulent}.

\begin{figure}
  \centerline{\includegraphics[width=0.88\textwidth]{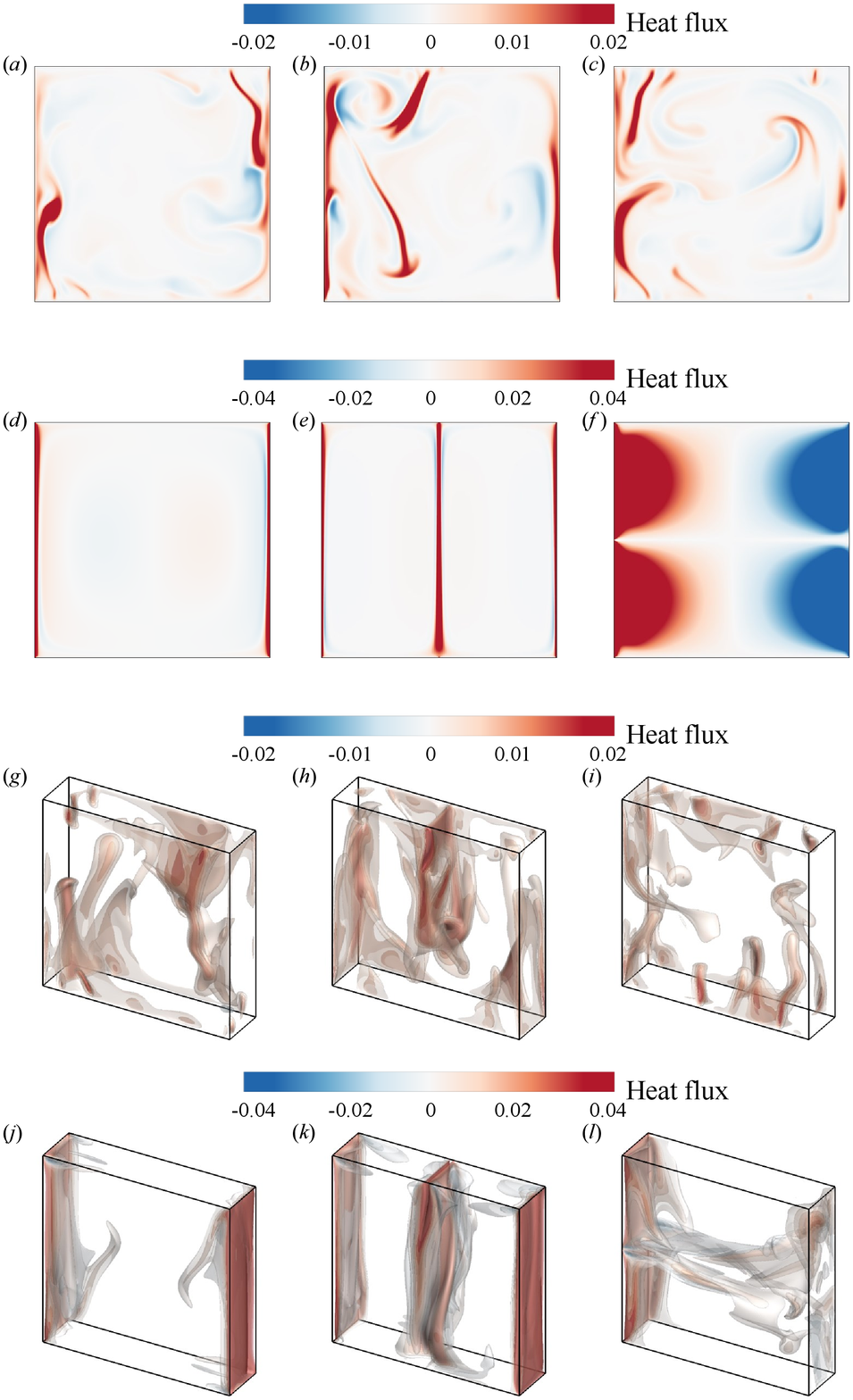}}
  \caption{Snapshots of vertical convective heat flux field at
  (\textit{a}-\textit{c}) $Re_{w} = 100$, (\textit{d}-\textit{f}) $Re_{w} = 4000$, (\textit{g}-\textit{i}) $Re_{w} = 100$ and $\Gamma_{\perp}=1/4$, (\textit{j}-\textit{l}) $Re_{w} = 3000$ and $\Gamma_{\perp}=1/4$,
  under (left column) the (1, 1) type, (middle-column) the (2, 1) type and (right-column) the (1, 2) type wall shear.}
\label{fig:Instant-Heat-Flux-Field}
\end{figure}
We now investigate how the local heat transfer properties are influenced by different types of wall shear.
In figure \ref{fig:Instant-Heat-Flux-Field}, we show the vertical convective heat flux field  $v \delta T$, where the temperature fluctuation is  $\delta T=T-(T_{hot}+T_{cold})/2$.
At small wall-shear strength $Re_{w}=100$ (see figures \ref{fig:Instant-Heat-Flux-Field}(\textit{a}-\textit{c}) for 2-D cases, and \ref{fig:Instant-Heat-Flux-Field}(\textit{g}-\textit{i}) for 3-D cases), the heat is transported mainly by the moving thermal plumes, and the magnitudes of vertical convective heat flux are relatively weak.
Under the (1, 1) type wall shear (see figures \ref{fig:Instant-Heat-Flux-Field}\textit{a},\textit{g}), plumes that carry heat mainly go up and down near the sidewalls;
under the (2, 1) type wall shear (see figures \ref{fig:Instant-Heat-Flux-Field}\textit{b},\textit{h}), plumes can also penetrate vertically in the bulk region of the cell, thus forming additional convection channels between the cold top wall and the hot bottom wall;
under the (1, 2) type wall shear (see figures \ref{fig:Instant-Heat-Flux-Field}\textit{c},\textit{i}), plumes that penetrate the bulk region of the cell exhibit horizontal motion at the mid-height of the cell.
At large wall-shear strength $Re_{w} = 4000$ (see figures \ref{fig:Instant-Heat-Flux-Field}(\textit{d}-\textit{f}) for 2-D cases) and $Re_{w}=3000$ (see figures \ref{fig:Instant-Heat-Flux-Field}(\textit{j}-\textit{l}) for 3-D cases), the vertical convective heat flux forms much more stable and regular convection channels, and their magnitudes are much stronger.
It should be noted that there are small regions of negative convective heat flux immediately adjacent to the regions of large positive convective heat flux, which is known as counter-gradient local heat transport \citep{gasteuil2007lagrangian,huang2013counter}.
The counter-gradient local heat transport essentially describes that both the LSC and the corner flows may contribute to heat transport in the 'wrong' direction:
hot (or cold) plumes can be brought back to the hot (or cold) plate by either the corner flows or the LSC.
The counter-gradient local heat transport is ubiquitous and can be found in 2-D and 3-D systems, either turbulent or laminar states.
An interesting finding is that under the (1, 2) type wall shear in the 2-D case (see figures \ref{fig:Instant-Heat-Flux-Field}\textit{f}), there exists strong negative vertical convective heat flux along the right vertical wall, which is opposite to the temperature gradient of the system.
Under the external wall shear, hot (or cold) fluids are forced to form a circulation in the bottom (or top) subregion of the cell.
When the hot (or cold) fluids fall (or rise) along the right vertical wall, they do not exchange heat with the other and do not lose their thermal energy at all, thus hot (or cold) fluids are swept back to the hot (or cold) walls and exhibit counter-gradient heat transport behaviour.
Previously, \cite{blass2020flow,blass2021effect} observed that by adding the Couette-type shear, the increase of heat transfer efficiency is due to elongated streaks generating vertical cross-stream motion,
while in our work, adding the $(m, n)$ type wall shear mainly facilitates a more coherent flow structure and forms more stable and stronger convection channels, particularly in the cases of laminar flows when the wall-shear strength is strong.

\begin{figure}
  \centerline{\includegraphics[width=0.5\textwidth]{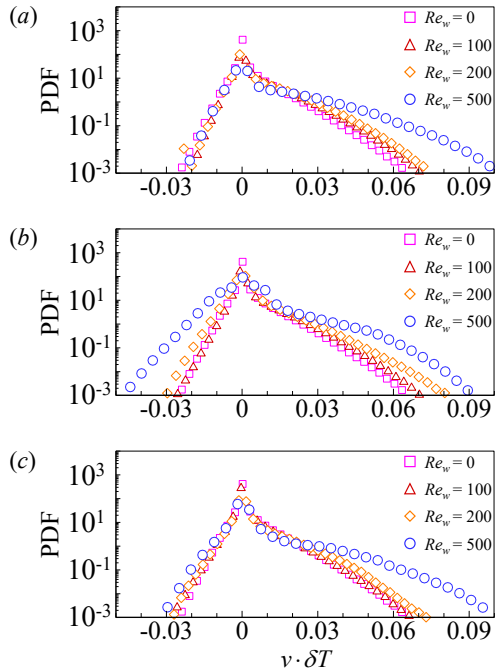}}
  \caption{The p.d.f.s of the heat flux measured in the whole cells for 2-D cases under (\textit{a}) the (1, 1) type wall shear, (\textit{b}) the (2, 1) type wall shear, and (\textit{c}) the (1, 2) type wall shear, when the flow is in turbulent state.}
\label{fig:pdfQy-wholeCell}
\end{figure}

In figure \ref{fig:pdfQy-wholeCell}, we further plot the p.d.f.s of the vertical convective heat flux $v \delta T$  in the whole cell for 2-D cases.
All the p.d.f.s have longer positive tails and shorter negative tails, implying strong upward convective heat transfer, yet there exists counter-gradient convective heat transfer \citep{huang2013counter}.
Under the wall shear, the strength of the upward convective heat transfer is enhanced with the increase of wall-shear strength in the whole cell;
meanwhile, we checked the p.d.f.s of convective heat flux in the bulk region (not shown here for clarity), and found that their shapes are much narrower, implying that heat exchange is weak in the bulk region, and hotter (or colder) fluids tend to flow upwards (or downwards) along the sidewalls.
Such strong counter-gradient convective heat transfer is consistent with our qualitative observation shown in figure \ref{fig:Instant-Heat-Flux-Field}.

\begin{figure}
  \centerline{\includegraphics[width=0.9\textwidth]{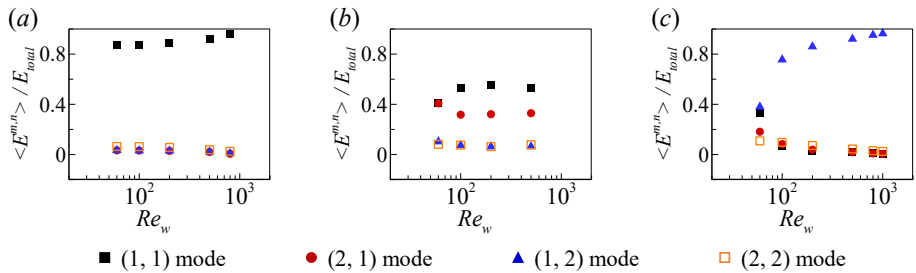}}
  \caption{Time-averaged energy contained in the first four Fourier modes as functions of $Re_{w}$ under (\textit{a}) the (1, 1) type wall shear, (\textit{b}) the (2, 1) type wall shear, (\textit{c}) the (1, 2) type wall shear.
   Note that the Fourier mode decomposition is  applied only when the flow is in a turbulent state for 2-D cases (refer to the phase diagram of flow states in figure \ref{fig:flowStates}).}
\label{fig:energyMode}
\end{figure}

In this work, we designed the $(m, n)$ type wall shear to facilitate the flow structure with $m$ rolls in the $x$-direction and $n$ rolls in the $y$-direction.
Under the imposed wall shear, to evaluate quantitatively whether the expected flow structure is dominated or not, we perform Fourier mode decomposition on the velocity field.
Fourier mode decomposition is a powerful tool to extract coherent structure in turbulent convection
\citep{petschel2011statistical,chandra2011dynamics,chandra2013flow,chong2018effect,wang2018flow}.
Specifically, the instantaneous velocity field $(u, v)$  is projected onto the Fourier basis $(\hat{u}^{m,n}, \hat{v}^{m,n})$  as
\begin{equation}
u(x,y,t)=\sum_{m,n} A_{x}^{m,n}(t)\hat{u}^{m,n}(x,y)
\end{equation}
\begin{equation}
v(x,y,t)=\sum_{m,n} A_{y}^{m,n}(t)\hat{v}^{m,n}(x,y)
\end{equation}
Here, the Fourier basis $(\hat{u}^{m,n}, \hat{v}^{m,n})$  is chosen as
\begin{equation}
\hat{u}^{m,n}(x,y)=2\sin(m\pi x)\cos(n\pi y)
\end{equation}
\begin{equation}
\hat{v}^{m,n}(x,y)=-2\cos(m\pi x)\sin(n\pi y)
\end{equation}
The instantaneous amplitude of the Fourier mode is then calculated as
\begin{equation}
A_{x}^{m,n}(t)=\langle u(x,y,t), \hat{u}^{m,n}(x,y) \rangle=\sum_{i}\sum_{j}u(x_{i},y_{i},t)\hat{u}^{m,n}(x_{i},y_{i})
\end{equation}
\begin{equation}
A_{y}^{m,n}(t)=\langle v(x,y,t), \hat{v}^{m,n}(x,y) \rangle=\sum_{i}\sum_{j}v(x_{i},y_{i},t)\hat{v}^{m,n}(x_{i},y_{i})
\end{equation}
where $\langle u, \hat{u} \rangle$  and $\langle v, \hat{v} \rangle$  denote the inner products of  $u$ and $\hat{u}$,  $v$ and $\hat{v}$, respectively.
The energy in each Fourier mode is calculated as  $E^{m,n}(t)=\sqrt{[A_{x}^{m,n}(t)]^{2}+[A_{y}^{m,n}(t)]^{2}}$, the total energy is calculated as  $E_{total}=\sum_{m,n}\langle E^{m,n} \rangle$, and  $\langle \cdot \rangle$ denotes the time average.
In figure \ref{fig:energyMode}, we plot the time-averaged energy as functions of $Re_{w}$ for various types of wall shear when the flow is in the turbulent state for 2-D cases.
Here, we consider  $m=1, 2$ and $n=1, 2$, namely the first four Fourier modes.
From figure \ref{fig:energyMode}(\textit{a}), we can see that under the (1, 1) type wall shear, the (1, 1) Fourier mode is indeed dominant.
Similarly, under the (1, 2) type wall shear, the (1, 2) Fourier mode is the dominant flow mode (see figure \ref{fig:energyMode}\textit{c}).
However, under the (2, 1) type wall shear, despite the energy percentage in the (2, 1) mode being much larger compared to that in the absence of wall shear, the (2, 1) mode does not contain the highest percentage of energy.
We can see from figure \ref{fig:energyMode}(\textit{b}) that the (1, 1) Fourier mode contains more energy than the expected (2, 1) mode.
To explain the discrepancy, we check the snapshots of the flow fields  and the heat flux fields (see figures \ref{fig:Instant-Flow-Field}(\textit{b}) and \ref{fig:Instant-Heat-Flux-Field}(\textit{b})), and observe that some hot (or cold) plumes lose their energy before reaching the cold top (or hot bottom) wall.
The plumes then fall (or turn back up) to form small rolls, thus substructures emerge inside the left-side big roll \citep{chen2019emergence}.
When the unstable small rolls inside the left-side big roll shrink their size, the Fourier mode decomposition that captures flows in the bulk region implies that the (1, 1) mode (i.e. one big roll in the whole cell) prevails.

\subsection{Stabilizing thermal turbulence via wall movement}
\begin{figure}
  \centerline{\includegraphics[width=0.85\textwidth]{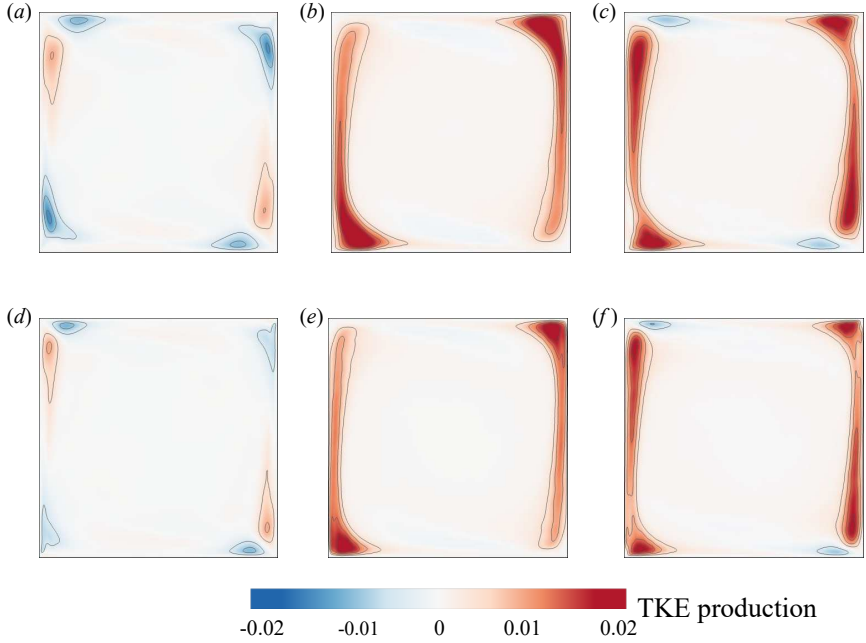}}
  \caption{(\textit{a},\textit{d}) The shear-produced turbulent kinetic energy (TKE),
  (\textit{b},\textit{e}) the buoyancy-produced TKE,
  and (\textit{c},\textit{f}) the total TKE production, under the (1, 1) type wall shear, for (\textit{a}-\textit{c}) $Re_{w} = 200$, (\textit{d}-\textit{f}) $Re_{w} = 500$.}
\label{fig:TKE}
\end{figure}

The original objective of imposing the $(m, n)$ type wall shear is to adjust the internal flow mode and control heat transfer properties,
while we found that by increasing the wall-shear strength, the thermal turbulence is relaminarized, and more surprisingly, the heat transfer efficiency of the convection cell in the laminar state is higher than that in the turbulent state.
In the previous subsection, we have explained that the enhancement of heat transfer efficiency at the laminar regime is due to the formation of more stable and stronger convection channels.
Below, we discuss further the origin of thermal turbulence laminarization.
We start by examining the turbulent kinetic energy (TKE) equation of incompressible thermal convection, which is written as
\begin{equation}
    \begin{split}
& \frac{\partial \mathcal{k}}{\partial t}+\overline{u_{j}}\partial_{j}\mathcal{k}
=-\overline{u_{i}'u_{j}'}\partial_{j} \overline{u_{i}} \\
& +\partial_{j}\left(-\overline{p'u_{j}'}+\sqrt{\frac{Pr}{Ra}}\partial_{j}\mathcal{k}-\frac{1}{2}\overline{u_{i}'u_{i}'u_{j}'} \right)
-\sqrt{\frac{Pr}{Ra}}\overline{\left(\partial_{j}u_{i}' \right)^{2}}+\overline{T'v'} \\
    \end{split}
\end{equation}
Here, the subscripts $i$ and $j$ are dummy indices,
$\mathcal{k}=\overline{u_{i}'u_{i}'}/2$ denotes the TKE, and the superscript ($'$) denotes the fluctuation part of an instantaneous flow variable.
The term $-\overline{u_{i}'u_{j}'}\partial_{j}\overline{u_{i}}$  represents shear-produced TKE, and the term $\overline{T'v'}$  represents buoyancy-produced TKE.
Because the flow remains laminar in a smaller range of $Re_{w}$ for 3-D cases, we discuss mainly the results for 2-D cases below.
In figure \ref{fig:TKE}, we show the shear-produced, buoyancy-produced and the total TKE production under the (1, 1) type wall shear as an example.
With increasing wall-shear strength, the shear-produced TKE is increasingly concentrated near the top left and bottom right corners of the convection cells (see figures \ref{fig:TKE}\textit{a},\textit{d}), where rising hot (or falling cold) plumes impact the cold (or hot) boundary layers.
Compared to the shear-produced TKE, the buoyancy-produced TKE is more intense (see figures \ref{fig:TKE}\textit{b},\textit{e}) and contributes a dominant part of the total TKE production (see figures \ref{fig:TKE}\textit{c},\textit{f}); meanwhile, with increasing wall-shear strength, the buoyancy-produced TKE becomes weaker.
Previously, in the absence of wall shear, \cite{xia2003particle} described quantitatively that the TKE  comes largely from the buoyant motions of thermal plumes based on the particle image velocimetry results.
With the aid of direct numerical simulation, $\overline{T'v'}$ is obtained directly in the whole convection cell, and we now provide direct evidence that thermal plumes are mainly responsible for the TKE production.

\begin{figure}
  \centerline{\includegraphics[width=0.75\textwidth]{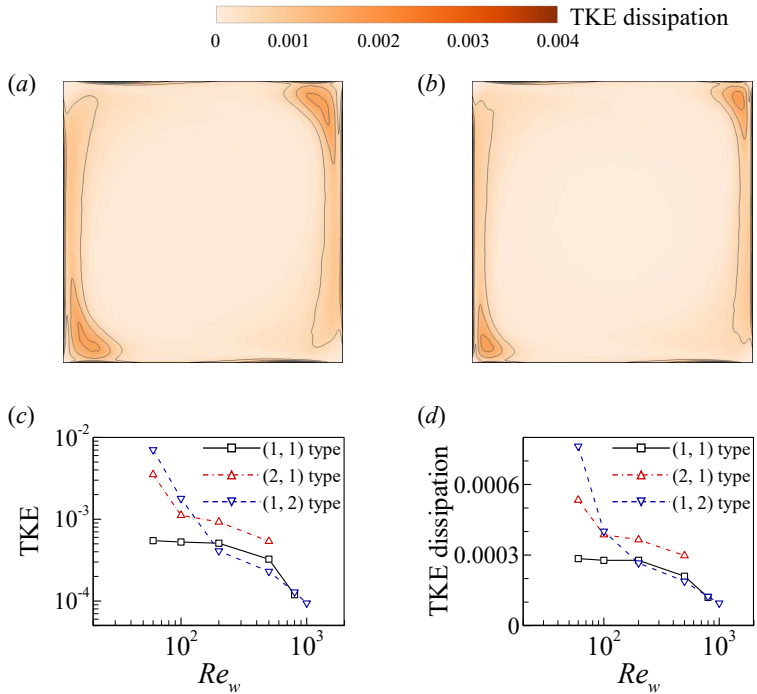}}
  \caption{The TKE dissipation for (\textit{a}) $Re_{w} = 200$ and (\textit{b}) $Re_{w} = 500$ under the (1, 1) type wall shear.
   (\textit{c}) The volume-averaged TKE, and (\textit{d}) the volume-averaged TKE dissipation, as functions of $Re_{w}$ for various types of wall shear.}
\label{fig:TKEdissipation}
\end{figure}

After analysing the TKE production, we now turn to the TKE dissipation.
In figures \ref{fig:TKEdissipation}(\textit{a}) and \ref{fig:TKEdissipation}(\textit{b}), we show the TKE dissipation under the (1, 1) type wall shear as an example.
We can see that intense TKE dissipation occurs in the top right and bottom left corners, namely, in the regions of plumes detachment.
Meanwhile, with increasing wall-shear strength, the TKE dissipation becomes weaker.
It should be noted that here we considered the dissipation term of  $\sqrt{Pr/Ra}\overline{(\partial_{j}u_{i}')^{2}}$ in the TKE equation, which is known as pseudo-dissipation by \cite{pope2000turbulent}.
Previously, \cite{zhang2017statistics} and \cite{bhattacharya2018complexity} analysed the statistics of TKE dissipation in terms of $\frac{1}{2}\sqrt{Pr/Ra}\overline{(\partial_{j}u_{i}'+\partial_{i}u_{j}')^{2}}$ in the canonical RB convection without wall shear.
We checked that the numerical differences between $\sqrt{Pr/Ra}\overline{(\partial_{j}u_{i}')^{2}}$ and $\frac{1}{2}\sqrt{Pr/Ra}\overline{(\partial_{j}u_{i}'+\partial_{i}u_{j}')^{2}}$ are indeed very small.
We then plot the volume-averaged TKE and the volume-averaged TKE dissipation as functions of $Re_{w}$ for various types of wall shear in figures \ref{fig:TKEdissipation}(\textit{c}) and \ref{fig:TKEdissipation}(\textit{d}).
With the increase of wall-shear strength, the volume-averaged TKE is indeed decreasing, eventually, the TKE vanishes, and the thermal turbulence is relaminarized.
However, the decreased TKE and the corresponding thermal turbulence laminarization are not caused by the viscous dissipation, and it is evident from figure \ref{fig:TKEdissipation}(\textit{d}) that the volume-averaged TKE dissipation is also decreasing.

\begin{figure}
  \centerline{\includegraphics[width=0.75\textwidth]{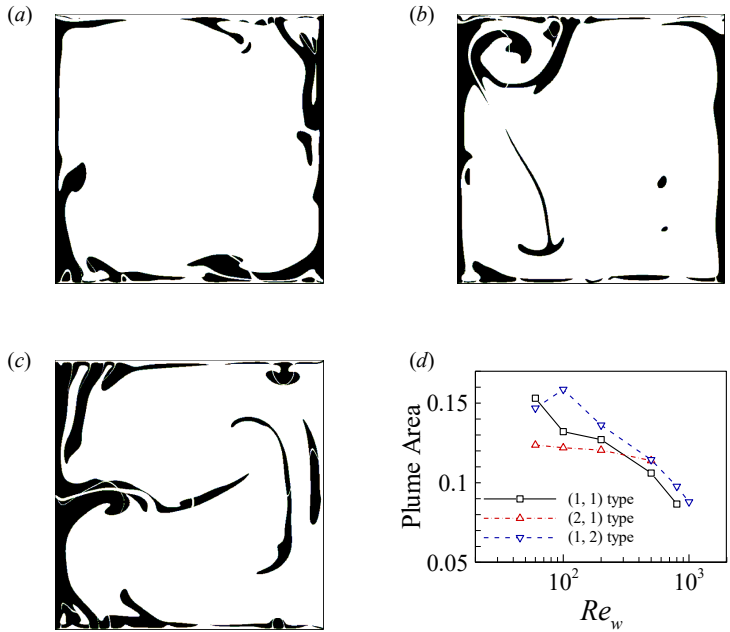}}
  \caption{Typical snapshots of plume field at $Re_{w} = 100$ for (\textit{a}) the (1, 1) type wall shear, (\textit{b}) the (2, 1) type wall shear, and (\textit{c}) the (1, 2) type wall shear. (\textit{d}) Time-averaged plume area in the cell as functions of $Re_{w}$ under the three types of wall shear.}
\label{fig:instant_PlumeField}
\end{figure}

The above analysis suggests that the plume plays a key role in thermal turbulence production and dissipation.
To identify the mechanism responsible for the thermal turbulence laminarization, we then analyse the spatial and temporal distributions of plumes.
In figures \ref{fig:instant_PlumeField}(\textit{a}-\textit{c}), we show typical snapshots of the instantaneous plume field under the three types of wall shear.
Here, the criteria to  identify thermal plumes quantitatively are similar to those used in \cite{huang2013confinement}, \cite{van2015plume} and \cite{zhang2017statistics}, namely
\begin{equation}
|T(\mathbf{x},t)-\langle T(\mathbf{x}) \rangle|>c\langle T_{rms}(\mathbf{x}) \rangle, \ \ \
\sqrt{Pr Ra}|v(\mathbf{x},t)T(\mathbf{x},t)|>cNu
\end{equation}
Here, $c$  is an empirical constant whose value can be chosen as  $0.8 \le c \le 1.2$, and we adopt the value $c=1$.
This criterion assumes that plumes occur in regions of local temperature maximum (or minimum), as well as regions where local convective heat flux is larger than the spatial and temporal averaged one.
We can see from figures \ref{fig:instant_PlumeField}(\textit{a}-\textit{c}) that this empirical criterion can extract the plume structures reasonably well in the sheared convection.
We also calculate the time-averaged plume area in the cell, and plot the plume areas as functions of $Re_{w}$.
From figure \ref{fig:instant_PlumeField}(\textit{d}), we can see that with the increase of wall-shear strength, plume areas generally decrease under all three types of wall shear.
Because thermal plumes are mainly responsible for TKE production, a reduced number of plumes indicates reduced TKE production.

We then examine the flow field during the laminarization process, as shown in figure \ref{fig:laminarization}, and the corresponding video can be viewed in supplementary movie 2.
Initially, an instantaneous flow field obtained at $Re_{w} = 200$ for the 2-D case, and $Re_{w} = 100$ for the 3-D case (i.e. the turbulent state), is used to start the simulation, in which the shear effects are relatively weak and the flow is buoyancy-dominated.
We can see plumes self-organize into the LSC, and large magnitudes of velocity vectors appear near the region where plumes erupt (see figure \ref{fig:laminarization}(\textit{a}) for the 2-D case).
When the wall-shear strength increases to $Re_{w} = 2000$ for the 2-D case, the plumes have less chance to detach from the boundary layers near the top and bottom walls, and they will be swept along the walls (see figure \ref{fig:laminarization}\textit{b}).
Because the organization of plume motions leads to the LSC in the turbulent RB convection cell \citep{xi2004laminar}, suppressing plume detachment will weaken the LSC.
In addition, hot (or cold) plumes are forced to sweep to the cold top (or hot bottom) wall (see figures \ref{fig:laminarization}\textit{c},\textit{d}), and thermal plumes exchange heat near the walls, while the temperature in the bulk region of the cell is more uniform and well-mixed (see figure \ref{fig:laminarization}\textit{e}).
Eventually, one regular big roll is formed, and hot and cold fluids flow along the wall, which is completely influenced by the external wall shear (see figure \ref{fig:laminarization}\textit{f}).
We can also see from figures \ref{fig:laminarization}(\textit{g}-\textit{l}) that the turbulence relaminarization process is similar for both 2-D and 3-D cases;
however, it is noteworthy that due to prominent shear instability effects in 3-D, the turbulence relaminarization is rare and occurs in a smaller range of wall-shear Reynolds number for 3-D cases.
\begin{figure}
  \centerline{\includegraphics[width=0.9\textwidth]{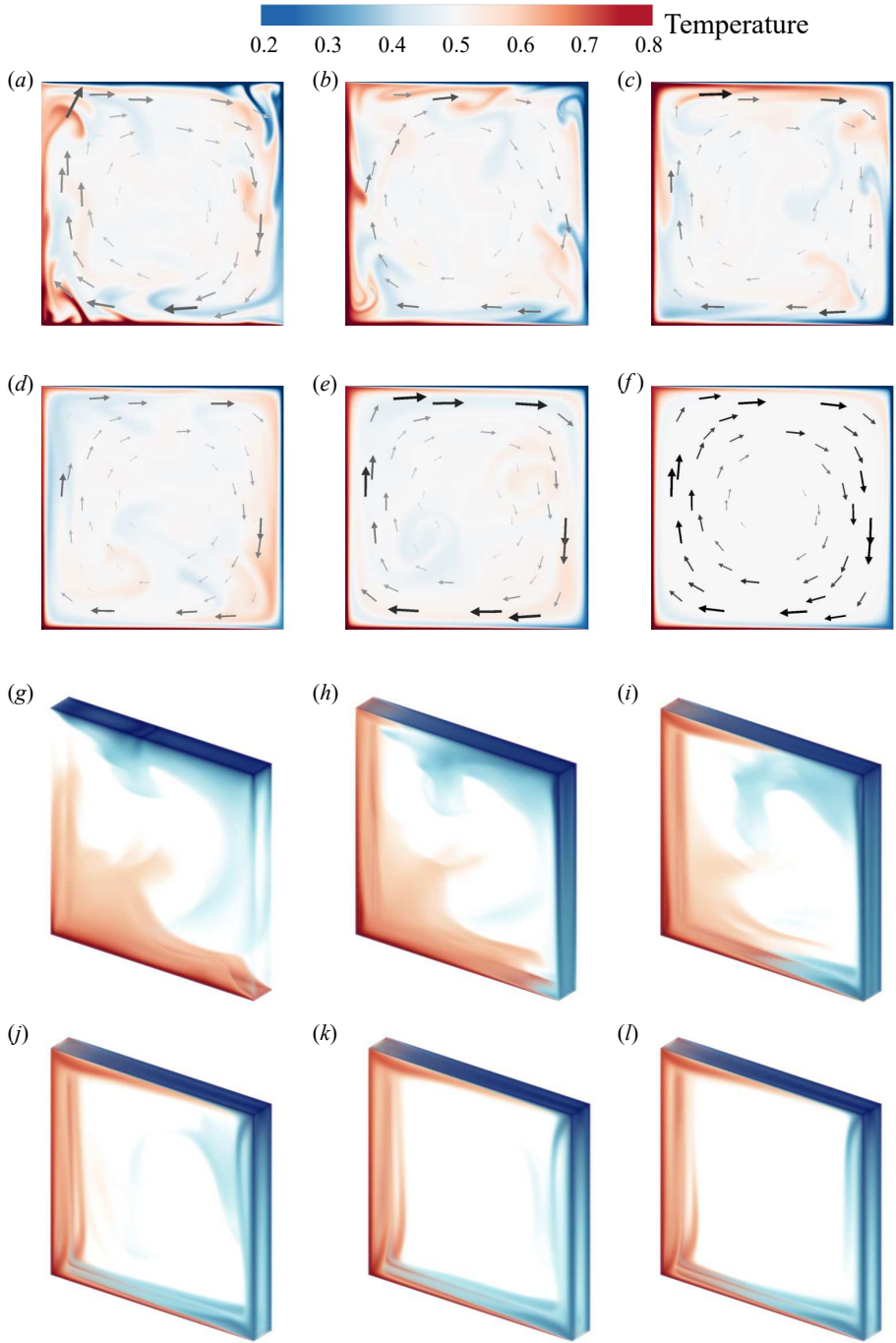}}
  \caption{Turbulence relaminarization process:
  time evolution of instantaneous flow fields (temperature contours and velocity vectors in 2-D, volume rendering of temperature field in 3-D).
  (\textit{a}-\textit{f}) Snapshots for the the 2-D case at $t = 0$, 2, 4, 8, 22 and 354  $t_{f}$, respectively, with wall-shear strength $Re_{w} = 2000$.
  (\textit{g}-\textit{l}) Snapshots for the 3-D case at $t = 0$, 2, 4, 8, 11 and 158  $t_{f}$, respectively, with $Re_{w} = 3000$.
  Initially, an instantaneous flow field obtained at $Re_{w} = 200$ for the 2-D case, and $Re_{w} = 100$ for the 3-D case, is used to start the simulation.}
\label{fig:laminarization}
\end{figure}

\subsection{Expenditure of mechanical energy due to external wall shear}

\begin{figure}
  \centerline{\includegraphics[width=0.45\textwidth]{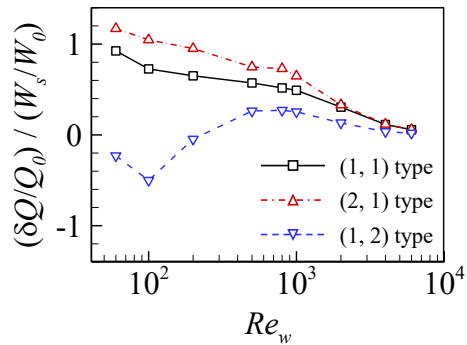}}
  \caption{The ratio between enhanced heat flux and imposed mechanical energy as a function of shear Reynolds number for various types of wall shear in the 2-D cases.}
\label{fig:energyEfficiency}
\end{figure}

We manipulated the internal flow modes via imposing external wall shear, and the corresponding heat transfer efficiency enhancement requires the expenditure of mechanical energy.
To evaluate whether such mechanical energy expenditure is worth it or not, we calculate the ratio between
the enhanced heat flux $\delta Q$ (which is further normalized by heat flux $Q_{0}$ in the absence of wall shear)
and the required mechanical energy $W_{s}$ due to wall shear (which is further normalized by energy dissipation due to viscosity $W_{0}$ in the absence of wall shear) as
\begin{equation}
\eta=\frac{\delta Q/Q_{0}}{W_{s}/W_{0}}
\end{equation}
Here, $\delta Q=Q_{\mathbf{u}_{w}}-Q_{0}$.
Generally, the heat flux $Q_{\mathbf{u}_{w}}$  is calculated as
\begin{equation}
Q_{\mathbf{u}_{w}}=\left \langle \int_{0}^{L}\left(-\kappa \frac{\partial T}{\partial \mathbf{n}} \right) dx  \right \rangle_{t}
\end{equation}
In the above, $\kappa$ denotes thermal conductivity of the fluids, and  $\langle \cdot \rangle_{t}$  denotes the time average.
To impose the wall shear, an additional external mechanical energy $W_{s}$ is required, which is calculated as
\begin{equation}
W_{s}=\left \langle  \oint_{\mathbf{l}} \left| \mu \frac{d\mathbf{u}_{w}}{d\mathbf{n}}\cdot \mathbf{u}_{w}  \right| d\mathbf{l}  \right \rangle_{t}
\end{equation}
Here, the integration $\oint_{\mathbf{l}}(\cdot)$ is performed along all the shear wall.
In the absence of wall shear, the energy dissipation due to viscosity in the convection cell is
\begin{equation}
W_{0}= \left \langle \int_{V} \frac{\mu}{2} \sum_{i,j}\left( \frac{\partial u_{i}}{\partial x_{j}}+\frac{\partial u_{j}}{\partial x_{i}} \right) dV \right \rangle_{t}
\end{equation}
The ratio between enhanced heat flux and imposed mechanical energy can be regarded as a metric that describes
the efficiency of facilitating heat transport via external shearing.
From figure \ref{fig:energyEfficiency}, we can see that for the (1, 1) and (2, 1) types of wall shear, the efficiency $\eta$ decreases monotonically with the increase of $Re_{w}$.
Recall that from figures \ref{fig:ReNu}(\textit{a}) and \ref{fig:ReNu}(\textit{c}), we found that $Nu$ increases monotonically with the increase of $Re_{w}$,
thus the enhanced $Nu$ requires a larger expenditure of mechanical energy at a larger $Re_{w}$.
For the (1, 2) type wall shear, the efficiency $\eta$ is negative at $Re_{w} \le 200$, which can be attributed to a shift in the flow structure from the LSC to the vertically stacked double-roll mode.
This transition in the flow mode results in a weakened $Nu$ within this range \citep{xu2020correlation}, as shown in figures \ref{fig:ReNu}(\textit{a}) and \ref{fig:ReNu}(\textit{c}).
At larger $Re_{w}$, the efficiency $\eta$ is positive, yet it does not exhibit monotonic behavior with the increase of $Re_{w}$.
Among the three types of wall shear, the (2, 1) type results in the highest efficiency $\eta$, because the flow is more coherent in the corresponding (2, 1) flow mode.
At the largest $Re_{w}$, all three types of wall shear exhibit very small values of efficiency $\eta$, implying that heat transfer enhancement comes at a very high price.
We deduce that with further increase in the wall-shear strength, the efficiency $\eta$ would approach the limit zero (but non-negative) value because the heat transfer increases monotonically at large $Re_{w}$.

\section{Conclusions}

In this work, we have performed direct numerical simulations of thermal convection under three different $(m, n)$ types of wall shear.
The $(m, n)$ type wall shear is imposed to facilitate $m$ rolls in the horizontal direction and $n$ rolls in the vertical direction.
Under the (1, 1) type, the (2, 1) type and the (1, 2) type wall shears, we can observe that the single-roll, the horizontally stacked double-roll, and the vertically stacked double-roll flow modes, respectively, are generally the prevailing flow modes in the convection cell.
With the increase of $Re_{w}$, we generally found enhanced heat transfer efficiency and global flow strength for all three types of wall shear.
However, even with the same magnitude of flow strength, the heat transfer efficiency of the convection cell varies significantly under different types of wall shear.
Specifically, the (2, 1) type wall shear results in the largest magnitude of heat transfer efficiency,
and the (1, 2) type wall shear results in the smallest,
which is consistent with our expectation that facilitating the horizontally stacked double-roll flow modes is efficient for heat transfer,
yet facilitating the vertically stacked double-roll is inefficient for heat transfer.

The original objective of imposing the wall shear was to manipulate flow mode to control heat transfer efficiency,
while it is found that by increasing the wall-shear strength, the thermal turbulence is relaminarized,
and more surprisingly, the heat transfer efficiency of the convection in the laminar state is higher than that in the turbulent state.
By examining the flow field and the convective heat flux field, we found that the enhancement of heat transfer efficiency at the laminar regime is due to the formation of more stable and stronger convection channels.

We explained the origin of thermal turbulence laminarization in the sheared convection cell.
Analysis of the shear-produced TKE (i.e. $-\overline{u_{i}'u_{j}'}\partial_{j}\overline{u_{i}}$) and the buoyancy-produced TKE (i.e. $\overline{T'v'}$) provides direct evidence that thermal plumes are mainly responsible for the TKE production.
We then quantitatively measured the changes in plume areas under the wall shear, and found that plumes are swept away by the wall shear once they are detached from the top cold and bottom hot walls,
and such a reduced number of thermal plumes decreases the TKE production in the bulk cell.

We evaluated whether the mechanical energy expenditure by wall shear is worth it or not.
We used the ratio between the enhanced heat flux and the required mechanical energy to quantify the efficiency of facilitating heat transport via external shearing.
We found that at a larger $Re_{w}$, although the heat transfer efficiency increases, it comes at a price of a larger expenditure of mechanical energy.

Finally, we emphasize that in mixed thermal convection, the heat transfer may not always monotonically increase with increasing shear.
For example, in the RB system with a Couette-type wall shear, \cite{blass2020flow,blass2021effect} found that with increasing wall shear, the heat transfer first decreases (due to the breakup of the thermal convection rolls) and then increases.
Another example is that in the RB system with an imposed constant horizontal pressure gradient, \cite{scagliarini2014heat,scagliarini2015law} found that with increasing longitudinal wind strength, the heat transfer also first decreases (due to plume sweeping) and then increases.
In our study, the heat transfer enhancement is a consequence of the moving adiabatic sidewalls advecting fluid in the vertical direction, thus facilitating the formation of stable and strong convection channels between the top cold wall and the bottom hot wall.

\section{Acknowledgments}
This work was supported by the National Natural Science Foundation of China (NSFC) through Grant Nos. 12272311 and 12125204,
and the 111 project of China (No. B17037).

\section{Declaration of interests}
The authors report no conflict of interest.

\section{Supplementary movies}
Supplementary movies are available at https://doi.org/10.1017/jfm.2023.173.

\section{Appendix A. Flow and heat transfer patterns in the canonical RB convection}
\begin{figure}
  \centerline{\includegraphics[width=\textwidth]{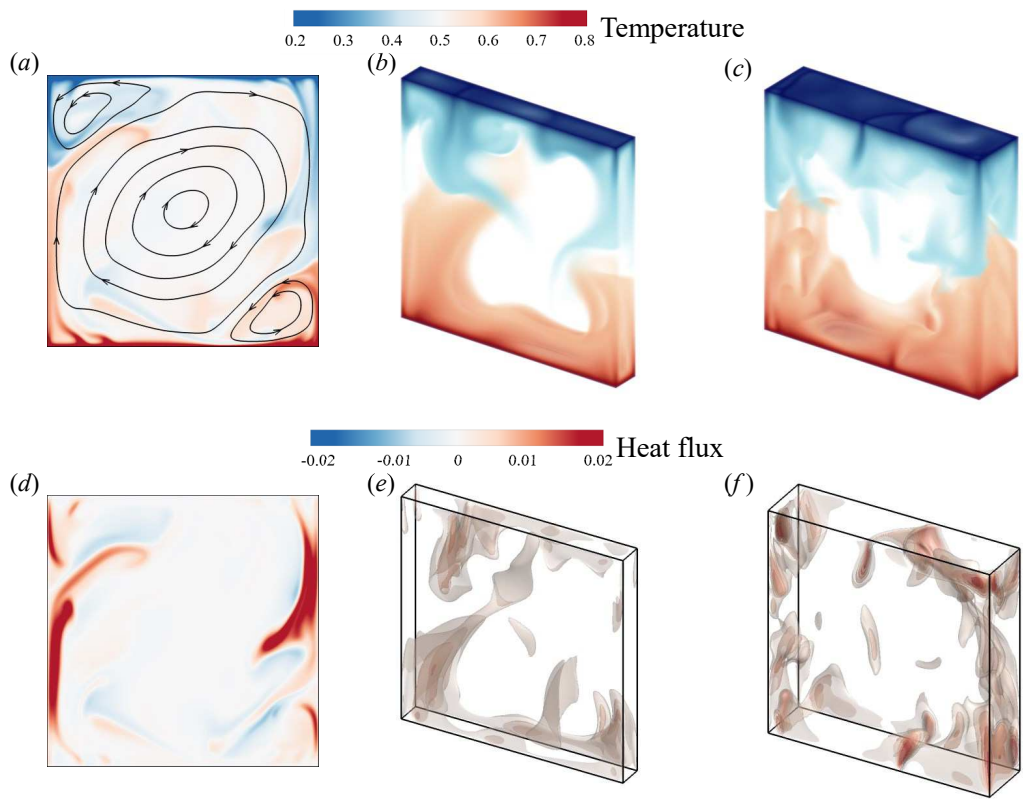}}
  \caption{Typical instantaneous (\textit{a}-\textit{c}) temperature field, and
  (\textit{d}-\textit{f}) vertical convective heat flux field, for the canonical RB convection at $Ra=10^{8}$ and $Pr=5.3$:
  (\textit{a},\textit{d}) 2-D case, (\textit{b},\textit{e}) 3-D case at $\Gamma_{\perp}=1/8$, (\textit{c},\textit{f}) 3-D case at $\Gamma_{\perp}=1/4$.
  }
\label{fig:Instant-withoutShear}
\end{figure}

In figure \ref{fig:Instant-withoutShear}, we show the typical instantaneous temperature and flow fields, as well as the vertical convective heat flux field for the canonical RB convection in the absence of wall shear.
In the 2-D case, we can see that there exists a well-defined LSC, together with counter-rotating corner rolls (figure \ref{fig:Instant-withoutShear}\textit{a}).
The LSC is in the form of a tilted ellipse sitting along a diagonal of the flow cell, with two secondary corner vortices that exist along the other diagonal.
Strong positive heat flux occurs in regions of rising hot (or falling cold) plumes (see figure \ref{fig:Instant-withoutShear}\textit{d}).
In the 3-D case, the very confined cell with $\Gamma_{\perp}=1/8$ exhibits similar flow and heat transfer patterns to those of the 2-D case, with persistent LSC (see figures \ref{fig:Instant-withoutShear}\textit{b},\textit{e}).
When the cell aspect ratio $\Gamma_{\perp}$ increases to $1/4$, the LSC is less stable and its shape becomes distorted (see figures \ref{fig:Instant-withoutShear}\textit{c},\textit{f}).

\section{Appendix B. Determination of flow states via time recordings and power spectral density}
\begin{figure}
  \centerline{\includegraphics[width=\textwidth]{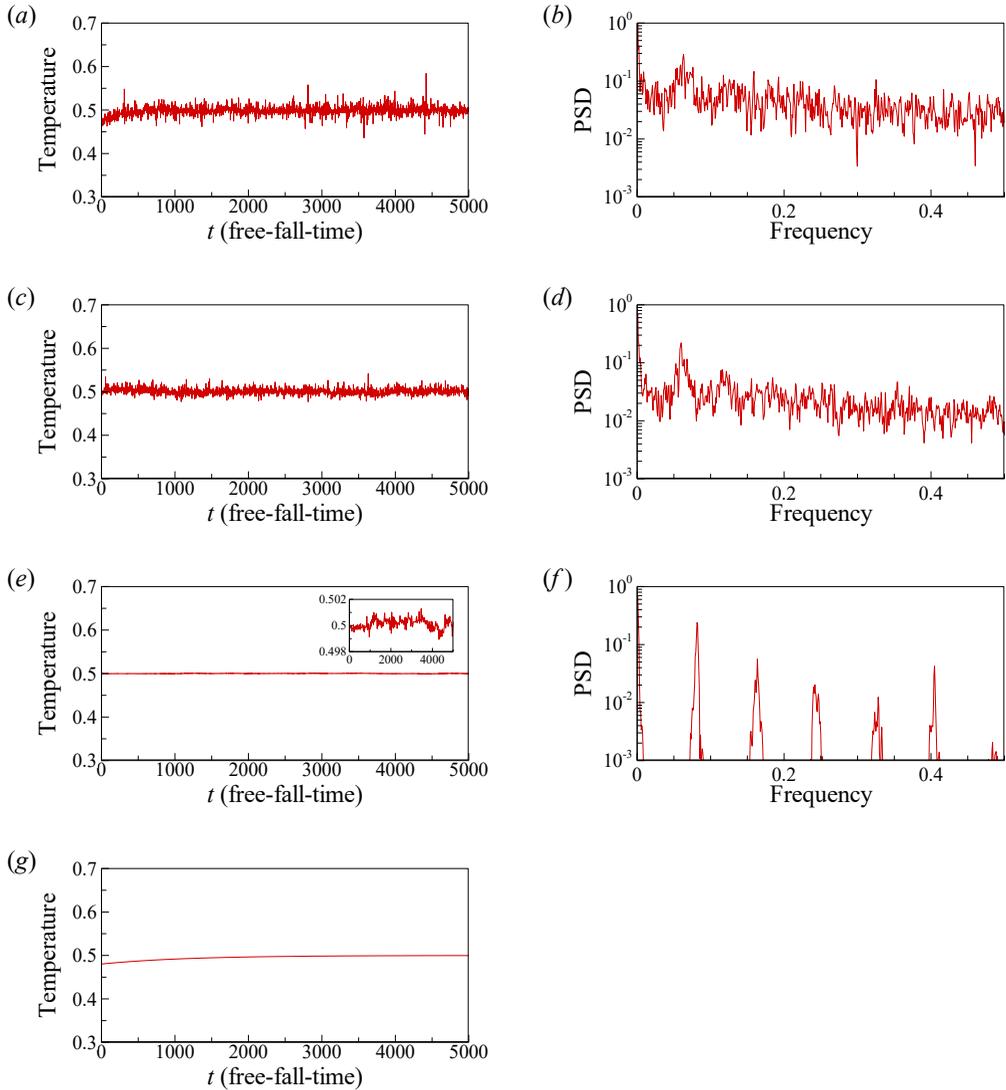}}
  \caption{(Left-column) Temperature series at the location (0.25, 0.5) in the 2-D convection cell under (1, 1) type wall shear.
  (Right-column) The PSD of the corresponding temperature series.
  Here, we have (\textit{a}, \textit{b}) $Re_{w} = 100$, (\textit{c}, \textit{d}) $Re_{w} = 500$, (\textit{e}, \textit{f}) $Re_{w} = 1000$, and (\textit{g}) $Re_{w} = 4000$.
  Note that the temperature series shown in (\textit{g}) eventually approaches a steady value, thus we did not calculate the corresponding PSD.}
\label{fig:PSD}
\end{figure}

In figure \ref{fig:PSD}, we give examples of temperature series at the location (0.25, 0.5) in the 2-D convection cell under (1, 1) type wall shear.
We also show the power spectral density (PSD) of the corresponding temperature series.
At $Re_{w} = 100$ and 500, the temperature fluctuates randomly around 0.5 (see figure \ref{fig:PSD}\textit{a},\textit{c}), and the corresponding PSD (see figure \ref{fig:PSD}\textit{b},\textit{d}) exhibit continuous spectra;
thus we determine the flow states as the turbulent state.
At $Re_{w} = 1000$, the fluctuation of the temperature series is within a smaller range (see figure \ref{fig:PSD}\textit{e} and its inset), and the corresponding PSD (see figure \ref{fig:PSD}\textit{f} exhibits characteristic peaks, suggesting that the flow is quasi-periodic;
thus we determine the flow states as the laminar state.
At $Re_{w} = 4000$, the temperature series gradually approaches a steady value of 0.5 (see figure \ref{fig:PSD}\textit{g}), thus we also determine the flow states as the laminar state.

\bibliographystyle{jfm}

\begin{thebibliography}{75}
\expandafter\ifx\csname natexlab\endcsname\relax\def\natexlab#1{#1}\fi
\def\au#1{#1} \def\ed#1{#1} \def\yr#1{#1}\def\at#1{#1}\def\jt#1{\textit{#1}}
  \def\bt#1{#1}\def\bvol#1{\textbf{#1}} \def\vol#1{#1} \def\pg#1{#1}
  \def\publ#1{#1}\def\arxiv#1{#1}\def\org#1{#1}\def\st#1{\textit{#1}}

\bibitem[Ahlers {\em et~al.\/}(2009)Ahlers, Grossmann \& Lohse]{ahlers2009heat}
{\sc \au{Ahlers, G.}, \au{Grossmann, S.} \& \au{Lohse, D.}} \yr{2009}  \at{Heat
  transfer and large scale dynamics in turbulent {R}ayleigh-{B}{\'e}nard
  convection}.  \jt{Rev. Mod. Phys.}  \bvol{81}~(2),  \pg{503--507}.

\bibitem[Batchelor(1959)]{batchelor1959small}
{\sc \au{Batchelor, G.~K.}} \yr{1959}  \at{Small-scale variation of convected
  quantities like temperature in turbulent fluid {P}art 1. {G}eneral discussion
  and the case of small conductivity}.  \jt{J. Fluid Mech.}  \bvol{5}~(1),
  \pg{113--133}.

\bibitem[Bhattacharya {\em et~al.\/}(2018)Bhattacharya, Pandey, Kumar \&
  Verma]{bhattacharya2018complexity}
{\sc \au{Bhattacharya, S.}, \au{Pandey, A.}, \au{Kumar, A.} \& \au{Verma,
  M.~K.}} \yr{2018}  \at{Complexity of viscous dissipation in turbulent thermal
  convection}.  \jt{Phys. Fluids}  \bvol{30}~(3),  \pg{031702}.

\bibitem[Blass {\em et~al.\/}(2021)Blass, Tabak, Verzicco, Stevens \&
  Lohse]{blass2021effect}
{\sc \au{Blass, A.}, \au{Tabak, P.}, \au{Verzicco, R.}, \au{Stevens, R. J.
  A.~M.} \& \au{Lohse, D.}} \yr{2021}  \at{The effect of {P}randtl number on
  turbulent sheared thermal convection}.  \jt{J. Fluid Mech.}  \bvol{910},
  \pg{A37}.

\bibitem[Blass {\em et~al.\/}(2020)Blass, Zhu, Verzicco, Lohse \&
  Stevens]{blass2020flow}
{\sc \au{Blass, A.}, \au{Zhu, X.-J.}, \au{Verzicco, R.}, \au{Lohse, D.} \&
  \au{Stevens, R. J. A.~M.}} \yr{2020}  \at{Flow organization and heat transfer
  in turbulent wall sheared thermal convection}.  \jt{J. Fluid Mech.}
  \bvol{897},  \pg{A22}.

\bibitem[Chandra \& Verma(2011)]{chandra2011dynamics}
{\sc \au{Chandra, M.} \& \au{Verma, M.~K.}} \yr{2011}  \at{Dynamics and
  symmetries of flow reversals in turbulent convection}.  \jt{Phys. Rev. E}
  \bvol{83}~(6),  \pg{067303}.

\bibitem[Chandra \& Verma(2013)]{chandra2013flow}
{\sc \au{Chandra, M.} \& \au{Verma, M.~K.}} \yr{2013}  \at{Flow reversals in
  turbulent convection via vortex reconnections}.  \jt{Phys. Rev. Lett.}
  \bvol{110}~(11),  \pg{114503}.

\bibitem[Chen {\em et~al.\/}(2019)Chen, Huang, Xia \& Xi]{chen2019emergence}
{\sc \au{Chen, X.}, \au{Huang, S.-D.}, \au{Xia, K.-Q.} \& \au{Xi, H.-D.}}
  \yr{2019}  \at{Emergence of substructures inside the large-scale circulation
  induces transition in flow reversals in turbulent thermal convection}.
  \jt{J. Fluid Mech.}  \bvol{877},  \pg{R1}.

\bibitem[Chill{\`a} {\em et~al.\/}(2004)Chill{\`a}, Rastello, Chaumat \&
  Castaing]{chilla2004long}
{\sc \au{Chill{\`a}, F.}, \au{Rastello, M.}, \au{Chaumat, S.} \& \au{Castaing,
  B.}} \yr{2004}  \at{Long relaxation times and tilt sensitivity in
  {R}ayleigh-{B}{\'e}nard turbulence}.  \jt{Eur. Phys. J. B}  \bvol{40}~(2),
  \pg{223--227}.

\bibitem[Chill{\`a} \& Schumacher(2012)]{chilla2012new}
{\sc \au{Chill{\`a}, F.} \& \au{Schumacher, J.}} \yr{2012}  \at{New
  perspectives in turbulent {R}ayleigh-{B}{\'e}nard convection}.  \jt{Eur.
  Phys. J. E}  \bvol{35}~(58),  \pg{1--25}.

\bibitem[Chong {\em et~al.\/}(2018)Chong, Wagner, Kaczorowski, Shishkina \&
  Xia]{chong2018effect}
{\sc \au{Chong, K.~L.}, \au{Wagner, S.}, \au{Kaczorowski, M.}, \au{Shishkina,
  O.} \& \au{Xia, K.-Q.}} \yr{2018}  \at{Effect of {P}randtl number on heat
  transport enhancement in {R}ayleigh-{B}{\'e}nard convection under geometrical
  confinement}.  \jt{Phys. Rev. Fluids}  \bvol{3}~(1),  \pg{013501}.

\bibitem[Chong {\em et~al.\/}(2017)Chong, Yang, Huang, Zhong, Stevens,
  Verzicco, Lohse \& Xia]{chong2017confined}
{\sc \au{Chong, K.~L.}, \au{Yang, Y.-T.}, \au{Huang, S.-D.}, \au{Zhong, J.-Q.},
  \au{Stevens, R. J. A.~M.}, \au{Verzicco, R.}, \au{Lohse, D.} \& \au{Xia,
  K.-Q.}} \yr{2017}  \at{Confined {R}ayleigh-{B}{\'e}nard, rotating
  {R}ayleigh-{B}{\'e}nard, and double diffusive convection: {A} unifying view
  on turbulent transport enhancement through coherent structure manipulation}.
  \jt{Phys. Rev. Lett.}  \bvol{119}~(6),  \pg{064501}.

\bibitem[Ciliberto {\em et~al.\/}(1996)Ciliberto, Cioni \&
  Laroche]{ciliberto1996large}
{\sc \au{Ciliberto, S.}, \au{Cioni, S.} \& \au{Laroche, C.}} \yr{1996}
  \at{Large-scale flow properties of turbulent thermal convection}.  \jt{Phys.
  Rev. E}  \bvol{54}~(6),  \pg{R5901}.

\bibitem[Ciliberto \& Laroche(1999)]{ciliberto1999random}
{\sc \au{Ciliberto, S.} \& \au{Laroche, C.}} \yr{1999}  \at{Random roughness of
  boundary increases the turbulent convection scaling exponent}.  \jt{Phys.
  Rev. Lett.}  \bvol{82}~(20),  \pg{3998}.

\bibitem[Deville {\em et~al.\/}(2002)Deville, Fischer \& Mund]{deville2002high}
{\sc \au{Deville, M.~O.}, \au{Fischer, P.~F.} \& \au{Mund, E.~H.}} \yr{2002}
  {\em High-order methods for incompressible fluid flow\/}, ,  \vol{vol.~9}.
  \publ{Cambridge University Press}.

\bibitem[Fischer(1997)]{fischer1997overlapping}
{\sc \au{Fischer, P.~F.}} \yr{1997}  \at{An overlapping {S}chwarz method for
  spectral element solution of the incompressible {N}avier--{S}tokes
  equations}.  \jt{J. Comput. Phys.}  \bvol{133}~(1),  \pg{84--101}.

\bibitem[Fischer {\em et~al.\/}(2002)Fischer, Kruse \&
  Loth]{fischer2002spectral}
{\sc \au{Fischer, P.~F.}, \au{Kruse, G.~W.} \& \au{Loth, F.}} \yr{2002}
  \at{Spectral element methods for transitional flows in complex geometries}.
  \jt{J. Sci. Comput.}  \bvol{17}~(1),  \pg{81--98}.

\bibitem[Gasteuil {\em et~al.\/}(2007)Gasteuil, Shew, Gibert, Chill{\`a},
  Castaing \& Pinton]{gasteuil2007lagrangian}
{\sc \au{Gasteuil, Y.}, \au{Shew, W.~L.}, \au{Gibert, M.}, \au{Chill{\`a}, F.},
  \au{Castaing, B.} \& \au{Pinton, J.-F.}} \yr{2007}  \at{{Lagrangian
  temperature, velocity, and local heat flux measurement in Rayleigh-B{\'e}nard
  convection}}.  \jt{Phys. Rev. Lett.}  \bvol{99}~(23),  \pg{234302}.

\bibitem[Guzman {\em et~al.\/}(2016)Guzman, Xie, Chen, Rivas, Sun, Lohse \&
  Ahlers]{guzman2016heat}
{\sc \au{Guzman, D.~N.}, \au{Xie, Y.-B.}, \au{Chen, S.-Y.}, \au{Rivas, D.~F.},
  \au{Sun, C.}, \au{Lohse, D.} \& \au{Ahlers, G.}} \yr{2016}  \at{Heat-flux
  enhancement by vapour-bubble nucleation in {R}ayleigh--{B}{\'e}nard
  turbulence}.  \jt{J. Fluid Mech.}  \bvol{787},  \pg{331--366}.

\bibitem[Gvozdi{\'c} {\em et~al.\/}(2018)Gvozdi{\'c}, Alm{\'e}ras, Mathai, Zhu,
  van Gils, Verzicco, Huisman, Sun \& Lohse]{gvozdic2018experimental}
{\sc \au{Gvozdi{\'c}, B.}, \au{Alm{\'e}ras, E.}, \au{Mathai, V.}, \au{Zhu,
  X.-J.}, \au{van Gils, D. P.~M.}, \au{Verzicco, R.}, \au{Huisman, S.~G.},
  \au{Sun, C.} \& \au{Lohse, D.}} \yr{2018}  \at{Experimental investigation of
  heat transport in homogeneous bubbly flow}.  \jt{J. Fluid Mech.}  \bvol{845},
   \pg{226--244}.

\bibitem[Heslot {\em et~al.\/}(1987)Heslot, Castaing \&
  Libchaber]{heslot1987transitions}
{\sc \au{Heslot, F.}, \au{Castaing, B.} \& \au{Libchaber, A.}} \yr{1987}
  \at{Transitions to turbulence in helium gas}.  \jt{Phys. Rev. A}
  \bvol{36}~(12),  \pg{5870}.

\bibitem[Huang {\em et~al.\/}(2013)Huang, Kaczorowski, Ni \&
  Xia]{huang2013confinement}
{\sc \au{Huang, S.-D.}, \au{Kaczorowski, M.}, \au{Ni, R.} \& \au{Xia, K.-Q.}}
  \yr{2013}  \at{Confinement-induced heat-transport enhancement in turbulent
  thermal convection}.  \jt{Phys. Rev. Lett.}  \bvol{111}~(10),  \pg{104501}.

\bibitem[Huang \& Xia(2016)]{huang2016effects}
{\sc \au{Huang, S.-D.} \& \au{Xia, K.-Q.}} \yr{2016}  \at{Effects of geometric
  confinement in quasi-2-{D} turbulent {R}ayleigh--{B}{\'e}nard convection}.
  \jt{J. Fluid Mech.}  \bvol{794},  \pg{639--654}.

\bibitem[Huang \& Zhou(2013)]{huang2013counter}
{\sc \au{Huang, Y.-X.} \& \au{Zhou, Q.}} \yr{2013}  \at{Counter-gradient heat
  transport in two-dimensional turbulent {R}ayleigh--{B}{\'e}nard convection}.
  \jt{J. Fluid Mech.}  \bvol{737},  \pg{R3}.

\bibitem[Hunt(1991)]{hunt1991industrial}
{\sc \au{Hunt, JCR}} \yr{1991}  \at{Industrial and environmental fluid
  mechanics}.  \jt{Annu. Rev. Fluid Mech.}  \bvol{23}~(1),  \pg{1--42}.

\bibitem[Jiang {\em et~al.\/}(2018)Jiang, Zhu, Mathai, Verzicco, Lohse \&
  Sun]{jiang2018controlling}
{\sc \au{Jiang, H.-C.}, \au{Zhu, X.-J.}, \au{Mathai, V.}, \au{Verzicco, R.},
  \au{Lohse, D.} \& \au{Sun, C.}} \yr{2018}  \at{Controlling heat transport and
  flow structures in thermal turbulence using ratchet surfaces}.  \jt{Phys.
  Rev. Lett.}  \bvol{120}~(4),  \pg{044501}.

\bibitem[Jin {\em et~al.\/}(2022)Jin, Wu, Zhang, Liu \& Zhou]{jin2022shear}
{\sc \au{Jin, T.-C.}, \au{Wu, J.-Z.}, \au{Zhang, Y.-Z.}, \au{Liu, Y.-L.} \&
  \au{Zhou, Q.}} \yr{2022}  \at{Shear-induced modulation on thermal convection
  over rough plates}.  \jt{J. Fluid Mech.}  \bvol{936},  \pg{A28}.

\bibitem[Kooij {\em et~al.\/}(2018)Kooij, Botchev, Frederix, Geurts, Horn,
  Lohse, van~der Poel, Shishkina, Stevens \& Verzicco]{kooij2018comparison}
{\sc \au{Kooij, G.~L.}, \au{Botchev, M.~A.}, \au{Frederix, E. M.~A.},
  \au{Geurts, B.~J.}, \au{Horn, S.}, \au{Lohse, D.}, \au{van~der Poel, E.~P.},
  \au{Shishkina, O.}, \au{Stevens, R. J. A.~M.} \& \au{Verzicco, R.}} \yr{2018}
   \at{Comparison of computational codes for direct numerical simulations of
  turbulent {R}ayleigh--{B}{\'e}nard convection}.  \jt{Comput. Fluids}
  \bvol{166},  \pg{1--8}.

\bibitem[Kraichnan(1962)]{kraichnan1962turbulent}
{\sc \au{Kraichnan, R.~H.}} \yr{1962}  \at{Turbulent thermal convection at
  arbitrary {P}randtl number}.  \jt{Phys. Fluids}  \bvol{5}~(11),
  \pg{1374--1389}.

\bibitem[K{\"u}hnen {\em et~al.\/}(2018)K{\"u}hnen, Song, Scarselli, Budanur,
  Riedl, Willis, Avila \& Hof]{kuhnen2018destabilizing}
{\sc \au{K{\"u}hnen, J.}, \au{Song, B.-F.}, \au{Scarselli, D.}, \au{Budanur,
  N.~B.}, \au{Riedl, M.}, \au{Willis, A.~P.}, \au{Avila, M.} \& \au{Hof, B.}}
  \yr{2018}  \at{Destabilizing turbulence in pipe flow}.  \jt{Nat. Phys.}
  \bvol{14}~(4),  \pg{386--390}.

\bibitem[Lakkaraju {\em et~al.\/}(2013)Lakkaraju, Stevens, Oresta, Verzicco,
  Lohse \& Prosperetti]{lakkaraju2013heat}
{\sc \au{Lakkaraju, R.}, \au{Stevens, R. J. A.~M.}, \au{Oresta, P.},
  \au{Verzicco, R.}, \au{Lohse, D.} \& \au{Prosperetti, A.}} \yr{2013}
  \at{Heat transport in bubbling turbulent convection}.  \jt{Proc. Natl. Acad.
  Sci.}  \bvol{110}~(23),  \pg{9237--9242}.

\bibitem[Liu \& Huisman(2020)]{liu2020heat}
{\sc \au{Liu, S.} \& \au{Huisman, S.~G.}} \yr{2020}  \at{Heat transfer
  enhancement in {R}ayleigh-{B}{\'e}nard convection using a single passive
  barrier}.  \jt{Phys. Rev. Fluids}  \bvol{5}~(12),  \pg{123502}.

\bibitem[Lohse \& Xia(2010)]{lohse2010small}
{\sc \au{Lohse, D.} \& \au{Xia, K.-Q.}} \yr{2010}  \at{Small-scale properties
  of turbulent {R}ayleigh-{B}{\'e}nard convection}.  \jt{Annu. Rev. Fluid
  Mech.}  \bvol{42},  \pg{335--364}.

\bibitem[Patera(1984)]{patera1984spectral}
{\sc \au{Patera, A.~T.}} \yr{1984}  \at{A spectral element method for fluid
  dynamics: laminar flow in a channel expansion}.  \jt{J. Comput. Phys.}
  \bvol{54}~(3),  \pg{468--488}.

\bibitem[Petschel {\em et~al.\/}(2011)Petschel, Wilczek, Breuer, Friedrich \&
  Hansen]{petschel2011statistical}
{\sc \au{Petschel, K.}, \au{Wilczek, M.}, \au{Breuer, M.}, \au{Friedrich, R.}
  \& \au{Hansen, U.}} \yr{2011}  \at{Statistical analysis of global wind
  dynamics in vigorous {R}ayleigh-{B}{\'e}nard convection}.  \jt{Phys. Rev. E}
  \bvol{84}~(2),  \pg{026309}.

\bibitem[van~der Poel {\em et~al.\/}(2014)van~der Poel, Ostilla-M{\'o}nico,
  Verzicco \& Lohse]{van2014effect}
{\sc \au{van~der Poel, E.~P.}, \au{Ostilla-M{\'o}nico, R.}, \au{Verzicco, R.}
  \& \au{Lohse, D.}} \yr{2014}  \at{Effect of velocity boundary conditions on
  the heat transfer and flow topology in two-dimensional
  {R}ayleigh-{B}{\'e}nard convection}.  \jt{Phys. Rev. E}  \bvol{90}~(1),
  \pg{013017}.

\bibitem[van~der Poel {\em et~al.\/}(2011)van~der Poel, Stevens \&
  Lohse]{van2011connecting}
{\sc \au{van~der Poel, E.~P.}, \au{Stevens, R. J. A.~M.} \& \au{Lohse, D.}}
  \yr{2011}  \at{Connecting flow structures and heat flux in turbulent
  {R}ayleigh-{B}{\'e}nard convection}.  \jt{Phys. Rev. E}  \bvol{84}~(4),
  \pg{045303}.

\bibitem[van~der Poel {\em et~al.\/}(2012)van~der Poel, Stevens, Sugiyama \&
  Lohse]{van2012flow}
{\sc \au{van~der Poel, E.~P.}, \au{Stevens, R. J. A.~M.}, \au{Sugiyama, K.} \&
  \au{Lohse, D.}} \yr{2012}  \at{Flow states in two-dimensional
  {R}ayleigh-{B}{\'e}nard convection as a function of aspect-ratio and
  {R}ayleigh number}.  \jt{Phys. Fluids}  \bvol{24}~(8),  \pg{085104}.

\bibitem[van~der Poel {\em et~al.\/}(2015)van~der Poel, Verzicco, Grossmann \&
  Lohse]{van2015plume}
{\sc \au{van~der Poel, E.~P.}, \au{Verzicco, R.}, \au{Grossmann, S.} \&
  \au{Lohse, D.}} \yr{2015}  \at{Plume emission statistics in turbulent
  {R}ayleigh--{B}{\'e}nard convection}.  \jt{J. Fluid Mech.}  \bvol{772},
  \pg{5--15}.

\bibitem[Pope(2000)]{pope2000turbulent}
{\sc \au{Pope, S.-B.}} \yr{2000} {\em Turbulent flows\/}.  \publ{Cambridge
  University Press}.

\bibitem[Roche {\em et~al.\/}(2002)Roche, Castaing, Chabaud \&
  H{\'e}bral]{roche2002prandtl}
{\sc \au{Roche, P.-E.}, \au{Castaing, B.}, \au{Chabaud, B.} \& \au{H{\'e}bral,
  B.}} \yr{2002}  \at{Prandtl and {R}ayleigh numbers dependences in
  {R}ayleigh-b{\'e}nard convection}.  \jt{EPL}  \bvol{58}~(5),  \pg{693}.

\bibitem[Rusaou{\"e}n {\em et~al.\/}(2018)Rusaou{\"e}n, Liot, Castaing, Salort
  \& Chill{\`a}]{rusaouen2018thermal}
{\sc \au{Rusaou{\"e}n, E.}, \au{Liot, O.}, \au{Castaing, B.}, \au{Salort, J.}
  \& \au{Chill{\`a}, F.}} \yr{2018}  \at{Thermal transfer in
  {R}ayleigh--{B}{\'e}nard cell with smooth or rough boundaries}.  \jt{J. Fluid
  Mech.}  \bvol{837},  \pg{443--460}.

\bibitem[Scagliarini {\em et~al.\/}(2015)Scagliarini, Einarsson, Gylfason \&
  Toschi]{scagliarini2015law}
{\sc \au{Scagliarini, A.}, \au{Einarsson, H.}, \au{Gylfason, {\'A}.} \&
  \au{Toschi, F.}} \yr{2015}  \at{Law of the wall in an unstably stratified
  turbulent channel flow}.  \jt{J. Fluid Mech.}  \bvol{781},  \pg{R5}.

\bibitem[Scagliarini {\em et~al.\/}(2014)Scagliarini, Gylfason \&
  Toschi]{scagliarini2014heat}
{\sc \au{Scagliarini, A.}, \au{Gylfason, {\'A}.} \& \au{Toschi, F.}} \yr{2014}
  \at{{Heat-flux scaling in turbulent Rayleigh-B{\'e}nard convection with an
  imposed longitudinal wind}}.  \jt{Phys. Rev. E}  \bvol{89}~(4),  \pg{043012}.

\bibitem[Scarselli {\em et~al.\/}(2019)Scarselli, K{\"u}hnen \&
  Hof]{scarselli2019relaminarising}
{\sc \au{Scarselli, D.}, \au{K{\"u}hnen, J.} \& \au{Hof, B.}} \yr{2019}
  \at{Relaminarising pipe flow by wall movement}.  \jt{J. Fluid Mech.}
  \bvol{867},  \pg{934--948}.

\bibitem[Shankar \& Deshpande(2000)]{shankar2000fluid}
{\sc \au{Shankar, P.~N.} \& \au{Deshpande, M.~D.}} \yr{2000}  \at{Fluid
  mechanics in the driven cavity}.  \jt{Annu. Rev. Fluid Mech.}  \bvol{32}~(1),
   \pg{93--136}.

\bibitem[Shraiman \& Siggia(1990)]{shraiman1990heat}
{\sc \au{Shraiman, B.~I.} \& \au{Siggia, E.~D.}} \yr{1990}  \at{Heat transport
  in high-{R}ayleigh-number convection}.  \jt{Phys. Rev. A}  \bvol{42}~(6),
  \pg{3650}.

\bibitem[Silano {\em et~al.\/}(2010)Silano, Sreenivasan \&
  Verzicco]{silano2010numerical}
{\sc \au{Silano, G.}, \au{Sreenivasan, K.~R.} \& \au{Verzicco, R.}} \yr{2010}
  \at{Numerical simulations of {R}ayleigh--{B}{\'e}nard convection for
  {P}randtl numbers between $10^{-1}$ and $10^{4}$ and {R}ayleigh numbers
  between $10^{5}$ and $10^{9}$}.  \jt{J. Fluid Mech.}  \bvol{662},
  \pg{409--446}.

\bibitem[Solomon \& Gollub(1990)]{solomon1990sheared}
{\sc \au{Solomon, T.~H.} \& \au{Gollub, J.~P.}} \yr{1990}  \at{{Sheared
  boundary layers in turbulent Rayleigh-B{\'e}nard convection}}.  \jt{Phys.
  Rev. Lett.}  \bvol{64}~(20),  \pg{2382}.

\bibitem[Stevens {\em et~al.\/}(2013)Stevens, Clercx \& Lohse]{stevens2013heat}
{\sc \au{Stevens, R. J. A.~M.}, \au{Clercx, H. J.~H.} \& \au{Lohse, D.}}
  \yr{2013}  \at{Heat transport and flow structure in rotating
  {R}ayleigh--{B}{\'e}nard convection}.  \jt{Eur. J. Mech. B-Fluids}
  \bvol{40},  \pg{41--49}.

\bibitem[Stevens {\em et~al.\/}(2009)Stevens, Zhong, Clercx, Ahlers \&
  Lohse]{stevens2009transitions}
{\sc \au{Stevens, R. J. A.~M.}, \au{Zhong, J.-Q.}, \au{Clercx, H. J.~H.},
  \au{Ahlers, G.} \& \au{Lohse, D.}} \yr{2009}  \at{Transitions between
  turbulent states in rotating {R}ayleigh-{B}{\'e}nard convection}.  \jt{Phys.
  Rev. Lett.}  \bvol{103}~(2),  \pg{024503}.

\bibitem[Sun {\em et~al.\/}(2005)Sun, Xi \& Xia]{sun2005azimuthal}
{\sc \au{Sun, C.}, \au{Xi, H.-D.} \& \au{Xia, K.-Q.}} \yr{2005}  \at{Azimuthal
  symmetry, flow dynamics, and heat transport in turbulent thermal convection
  in a cylinder with an aspect ratio of 0.5}.  \jt{Phys. Rev. Lett.}
  \bvol{95}~(7),  \pg{074502}.

\bibitem[Wagner \& Shishkina(2015)]{wagner2015heat}
{\sc \au{Wagner, S.} \& \au{Shishkina, O.}} \yr{2015}  \at{Heat flux
  enhancement by regular surface roughness in turbulent thermal convection}.
  \jt{J. Fluid Mech.}  \bvol{763},  \pg{109--135}.

\bibitem[Wang {\em et~al.\/}(2020)Wang, Zhou \& Sun]{wang2020vibration}
{\sc \au{Wang, B.-F.}, \au{Zhou, Q.} \& \au{Sun, C.}} \yr{2020}
  \at{Vibration-induced boundary-layer destabilization achieves massive
  heat-transport enhancement}.  \jt{Sci. Adv.}  \bvol{6}~(21),  \pg{eaaz8239}.

\bibitem[Wang {\em et~al.\/}(2018)Wang, Xia, Wang, Sun, Zhou \&
  Wan]{wang2018flow}
{\sc \au{Wang, Q.}, \au{Xia, S.-N.}, \au{Wang, B.-F.}, \au{Sun, D.-J.},
  \au{Zhou, Q.} \& \au{Wan, Z.-H.}} \yr{2018}  \at{Flow reversals in
  two-dimensional thermal convection in tilted cells}.  \jt{J. Fluid Mech.}
  \bvol{849},  \pg{355--372}.

\bibitem[Wang {\em et~al.\/}(2019)Wang, Mathai \& Sun]{wang2019self}
{\sc \au{Wang, Z.-Q.}, \au{Mathai, V.} \& \au{Sun, C.}} \yr{2019}
  \at{Self-sustained biphasic catalytic particle turbulence}.  \jt{Nat.
  Commun.}  \bvol{10}~(1),  \pg{1--7}.

\bibitem[Weiss \& Ahlers(2011)]{weiss2011turbulent}
{\sc \au{Weiss, S.} \& \au{Ahlers, G.}} \yr{2011}  \at{Turbulent
  {R}ayleigh--{B}{\'e}nard convection in a cylindrical container with aspect
  ratio $\gamma$= 0.50 and {P}randtl number {P}r=4.38}.  \jt{J. Fluid Mech.}
  \bvol{676},  \pg{5--40}.

\bibitem[Xi {\em et~al.\/}(2004)Xi, Lam \& Xia]{xi2004laminar}
{\sc \au{Xi, H.-D.}, \au{Lam, S.} \& \au{Xia, K.-Q.}} \yr{2004}  \at{From
  laminar plumes to organized flows: the onset of large-scale circulation in
  turbulent thermal convection}.  \jt{J. Fluid Mech.}  \bvol{503},
  \pg{47--56}.

\bibitem[Xi \& Xia(2008)]{xi2008flow}
{\sc \au{Xi, H.-D.} \& \au{Xia, K.-Q.}} \yr{2008}  \at{Flow mode transitions in
  turbulent thermal convection}.  \jt{Phys. Fluids}  \bvol{20}~(5),
  \pg{055104}.

\bibitem[Xi {\em et~al.\/}(2016)Xi, Zhang, Hao \& Xia]{xi2016higher}
{\sc \au{Xi, H.-D.}, \au{Zhang, Y.-B.}, \au{Hao, J.-T.} \& \au{Xia, K.-Q.}}
  \yr{2016}  \at{Higher-order flow modes in turbulent {R}ayleigh--{B}{\'e}nard
  convection}.  \jt{J. Fluid Mech.}  \bvol{805},  \pg{31--51}.

\bibitem[Xia(2013)]{xia2013current}
{\sc \au{Xia, K.-Q.}} \yr{2013}  \at{Current trends and future directions in
  turbulent thermal convection}.  \jt{Theor. Appl. Mech. Lett.}  \bvol{3}~(5),
  \pg{052001}.

\bibitem[Xia \& Lui(1997)]{xia1997turbulent}
{\sc \au{Xia, K.-Q.} \& \au{Lui, S.-L.}} \yr{1997}  \at{Turbulent thermal
  convection with an obstructed sidewall}.  \jt{Phys. Rev. Lett.}
  \bvol{79}~(25),  \pg{5006}.

\bibitem[Xia {\em et~al.\/}(2003)Xia, Sun \& Zhou]{xia2003particle}
{\sc \au{Xia, K.-Q.}, \au{Sun, C.} \& \au{Zhou, S.-Q.}} \yr{2003}  \at{Particle
  image velocimetry measurement of the velocity field in turbulent thermal
  convection}.  \jt{Phys. Rev. E}  \bvol{68}~(6),  \pg{066303}.

\bibitem[Xu {\em et~al.\/}(2020)Xu, Chen, Wang \& Xi]{xu2020correlation}
{\sc \au{Xu, A.}, \au{Chen, X.}, \au{Wang, F.} \& \au{Xi, H.-D.}} \yr{2020}
  \at{Correlation of internal flow structure with heat transfer efficiency in
  turbulent {R}ayleigh--{B}{\'e}nard convection}.  \jt{Phys. Fluids}
  \bvol{32}~(10),  \pg{105112}.

\bibitem[Xu {\em et~al.\/}(2021)Xu, Chen \& Xi]{xu2021tristable}
{\sc \au{Xu, A.}, \au{Chen, X.} \& \au{Xi, H.-D.}} \yr{2021}  \at{Tristable
  flow states and reversal of the large-scale circulation in two-dimensional
  circular convection cells}.  \jt{J. Fluid Mech.}  \bvol{910},  \pg{A33}.

\bibitem[Xu \& Li(2023)]{xu2023multi}
{\sc \au{Xu, A.} \& \au{Li, B.-T.}} \yr{2023}  \at{{Multi-GPU thermal lattice
  Boltzmann simulations using OpenACC and MPI}}.  \jt{Int. J. Heat Mass
  Transf.}  \bvol{201},  \pg{123649}.

\bibitem[Xu {\em et~al.\/}(2019)Xu, Shi \& Xi]{xu2019lattice}
{\sc \au{Xu, A.}, \au{Shi, L.} \& \au{Xi, H.-D.}} \yr{2019}  \at{{Lattice
  Boltzmann simulations of three-dimensional thermal convective flows at high
  Rayleigh number}}.  \jt{Int. J. Heat Mass Transf.}  \bvol{140},
  \pg{359--370}.

\bibitem[Xu {\em et~al.\/}(2017)Xu, Shi \& Zhao]{xu2017accelerated}
{\sc \au{Xu, A.}, \au{Shi, L.} \& \au{Zhao, T.S.}} \yr{2017}  \at{{Accelerated
  lattice Boltzmann simulation using GPU and OpenACC with data management}}.
  \jt{Int. J. Heat Mass Transf.}  \bvol{109},  \pg{577--588}.

\bibitem[Yang {\em et~al.\/}(2020{\natexlab{{\em a\/}}})Yang, Chong, Wang,
  Verzicco, Shishkina \& Lohse]{yang2020periodically}
{\sc \au{Yang, R.}, \au{Chong, K.~L.}, \au{Wang, Q.}, \au{Verzicco, R.},
  \au{Shishkina, O.} \& \au{Lohse, D.}} \yr{2020{\natexlab{{\em a\/}}}}
  \at{Periodically modulated thermal convection}.  \jt{Phys. Rev. Lett.}
  \bvol{125}~(15),  \pg{154502}.

\bibitem[Yang {\em et~al.\/}(2022)Yang, Zhang, Wang, Dong \&
  Zhou]{yang2022dynamic}
{\sc \au{Yang, W.-W.}, \au{Zhang, Y.-Z.}, \au{Wang, B.-F.}, \au{Dong, Y.-H.} \&
  \au{Zhou, Q.}} \yr{2022}  \at{Dynamic coupling between carrier and dispersed
  phases in {R}ayleigh--{B}{\'e}nard convection laden with inertial isothermal
  particles}.  \jt{J. Fluid Mech.}  \bvol{930},  \pg{A24}.

\bibitem[Yang {\em et~al.\/}(2020{\natexlab{{\em b\/}}})Yang, Verzicco, Lohse
  \& Stevens]{yang2020rotation}
{\sc \au{Yang, Y.-T.}, \au{Verzicco, R.}, \au{Lohse, D.} \& \au{Stevens, R. J.
  A.~M.}} \yr{2020{\natexlab{{\em b\/}}}}  \at{What rotation rate maximizes
  heat transport in rotating {R}ayleigh-{B}{\'e}nard convection with {P}randtl
  number larger than one?}  \jt{Phys. Rev. Fluids}  \bvol{5}~(5),  \pg{053501}.

\bibitem[Zhang {\em et~al.\/}(2022)Zhang, Dong \& Xia]{zhang2022exploring}
{\sc \au{Zhang, L.}, \au{Dong, J.} \& \au{Xia, K.-Q.}} \yr{2022}  \at{Exploring
  the plume and shear effects in turbulent {R}ayleigh--{B}{\'e}nard convection
  with effective horizontal buoyancy under streamwise and spanwise geometrical
  confinements}.  \jt{J. Fluid Mech.}  \bvol{940},  \pg{A37}.

\bibitem[Zhang {\em et~al.\/}(2017)Zhang, Zhou \& Sun]{zhang2017statistics}
{\sc \au{Zhang, Y.}, \au{Zhou, Q.} \& \au{Sun, C.}} \yr{2017}  \at{Statistics
  of kinetic and thermal energy dissipation rates in two-dimensional turbulent
  {R}ayleigh--{B}{\'e}nard convection}.  \jt{J. Fluid Mech.}  \bvol{814},
  \pg{165--184}.

\bibitem[Zhong {\em et~al.\/}(2009)Zhong, Stevens, Clercx, Verzicco, Lohse \&
  Ahlers]{zhong2009prandtl}
{\sc \au{Zhong, J.-Q.}, \au{Stevens, R. J. A.~M.}, \au{Clercx, H. J.~H.},
  \au{Verzicco, R.}, \au{Lohse, D.} \& \au{Ahlers, G.}} \yr{2009}
  \at{Prandtl-, {R}ayleigh-, and {R}ossby-number dependence of heat transport
  in turbulent rotating {R}ayleigh-{B}{\'e}nard convection}.  \jt{Phys. Rev.
  Lett.}  \bvol{102}~(4),  \pg{044502}.

\bibitem[Zhu {\em et~al.\/}(2019)Zhu, Stevens, Shishkina, Verzicco \&
  Lohse]{zhu2019scaling}
{\sc \au{Zhu, X.-J.}, \au{Stevens, R. J. A.~M.}, \au{Shishkina, O.},
  \au{Verzicco, R.} \& \au{Lohse, D.}} \yr{2019}  \at{Scaling enabled by
  multiscale wall roughness in {R}ayleigh--{B}{\'e}nard turbulence}.  \jt{J.
  Fluid Mech.}  \bvol{869},  \pg{R4}.

\end{thebibliography}

\end{document}